%% file: V2V_final.tex
\documentclass[10pt,comsoc]{IEEEtran}

\input{input.tex}

\usepackage{epstopdf}
\usepackage{enumerate}
\usepackage{algorithmicx}
\usepackage{algorithm}
\usepackage{amsmath}
\usepackage[noend]{algpseudocode}
\usepackage{float}
\usepackage{hyperref}
\usepackage{color}
\usepackage{makeidx}
\usepackage{bbm}
\usepackage{graphicx}
\usepackage{booktabs}
\usepackage{subfigure}
\usepackage{multirow}

\begin{document}
\title{DeepSense-V2V: A Vehicle-to-Vehicle Multi-Modal Sensing, Localization, and Communications Dataset}

\author{João Morais, Gouranga Charan, Nikhil Srinivas, and Ahmed Alkhateeb \thanks{The authors are with the School of Electrical, Computer, and Energy Engineering, Arizona State University. Emails: \{joao, gcharan, tvsrini1,  alkhateeb\}@asu.edu. This work is supported by the National Science Foundation under Grant No. 2048021.}}

\maketitle

\begin{abstract}
High data-rate and low-latency vehicle-to-vehicle (V2V) communication is essential for future intelligent transport systems to enable coordination, enhance safety, and support distributed computing and intelligence requirements. Developing effective communication strategies, however, demands realistic test scenarios and datasets. This is important at the high frequency bands where more spectrum is available yet harvesting this bandwidth is challenged by the need for direction transmission and the sensitivity of signal propagation to blockages. This work presents the first large-scale multi-modal dataset for studying mmWave vehicle-to-vehicle communications. It presents a two-vehicle testbed that comprises data from a 360º camera, four radars, four 60 GHz phased arrays, a 3D lidar, and two precise GPSs. The dataset contains vehicles driving during the day and night for 120 km in intercity and rural settings, with speeds up to 100 km per hour. More than one million objects were detected across all images, from trucks to bicycles. This work further includes detailed dataset statistics that prove the coverage of various situations and highlight how this dataset can enable novel machine-learning applications.
\end{abstract}

\vspace{-.02cm}

\section{Introduction} \label{sec:Intro}

Vehicle-to-vehicle (V2V) communication has become increasingly essential in intelligent transportation systems (ITS) for enabling vehicles to exchange critical information, enhancing safety, traffic efficiency, and the overall driving experience \cite{harding2014vehicle}. However, the current methods of V2V communication face challenges with the increasing volume and complexity of data being exchanged, which might limit the effectiveness of the ITS \cite{zeadally2020tutorial}. This demand for higher data rates in V2V communication motivates the exploration of higher frequency bands such as millimeter wave (mmWave) and sub-terahertz (sub-THz) frequencies. The mmWave/sub-THz frequency ranges offer larger bandwidths, making them well-suited for supporting the high-speed and data-intensive requirements of V2V communication systems \cite{rappaport2019wireless}. Additionally, the availability of large antenna arrays and beamforming capabilities in mmWave/sub-THz V2V communication systems enable robust and efficient communication, mitigating the effects of interference and signal attenuation in dynamic and congested environments. Adopting advanced wireless communication technologies in V2V systems facilitates reliable data exchange between vehicles, even in high-speed scenarios, where rapid and accurate information dissemination is crucial for collision avoidance, cooperative driving, and other vehicular applications.

Further, future wireless systems, specifically in 6G and beyond, are envisioned to incorporate communication, multi-modal sensing, and positioning capabilities as integral components \cite{alkhateeb2023real, viswanathan2020communications}. These systems are anticipated to implement co-existing communication and sensing functionalities or leverage one to enhance the other, accentuating the growing importance of the synergy between multi-modal sensing and communication. This synergy has been driving key research directions such as multi-modal sensing-aided communication \cite{charan2022vision, charan2021vision, morais2022position, charan2022multi, jiang2022lidar, demirhan2022radar, charan2022drone, wu2023proactively, charan2022blockage} and integrated sensing and communication \cite{demirhan2023integrated}. Moreover, with the rise of autonomous vehicles, there is an increasing focus on equipping vehicles with multiple sensors, such as radar, LiDAR, and cameras, enabling vehicles to gather comprehensive situational awareness. Incorporating co-located communication and sensing functionalities will likely be the key to enabling reliable and efficient V2V communication. Multi-modal sensing capabilities can help navigate complex and dynamic scenarios on the road effectively. A detailed perception of the environment can enhance V2V communication reliability, facilitate advanced decision-making algorithms, and improve overall safety and efficiency in complex and dynamic environments. Despite these benefits, fully realizing efficient V2V communication presents challenges, particularly when dealing with mmWave/sub-THz frequency communication.

The realization of efficient mmWave vehicle-to-vehicle communication benefits from (i) the development of sophisticated detection and tracking algorithms and (ii) the resolution of the unique challenges posed by mmWave/sub-THz communication systems. First, the development of sophisticated detection and tracking algorithms can support directional beamforming and blockage detection/tracking in mmWave systems.  Second, the utilization of mmWave/sub-THz frequencies introduces challenges. For instance, adjusting the narrow beams in these communication systems with large antenna arrays is typically associated with large training overhead that scales with the number of antennas, making it challenging to support high-mobility applications such as V2V communication. Further, line-of-sight (LOS)link blockages such as buildings and other vehicles can disrupt communication and challenge the link reliability. Although several multi-modal datasets \cite{ caesar2020nuscenes, llc2019waymo, pham20203d} recently have been made available targeting autonomous vehicles, the development of large-scale datasets designed explicitly for V2V communication is lacking. To address these challenges, it is crucial to create comprehensive multi-modal sensing-aided V2V communication datasets that capture real-world scenarios, enabling researchers to design and evaluate algorithms and protocols for this specific context. 

\begin{table*}[!t]
	\caption{Comparative summary of key characteristics of state-of-the-art vehicular datasets.}
	\centering
	\setlength{\tabcolsep}{5pt}
	\renewcommand{\arraystretch}{1.2}
	\resizebox{\textwidth}{!}{
	\begin{tabular}{|c|c|c|c|c|c|c|c|c|c|c|}
	\hline
	Dataset               & Year          & Application                                                                     & \begin{tabular}[c]{@{}c@{}}Wireless \\ Comm.\end{tabular} & Scenes    & Size (hr) & \begin{tabular}[c]{@{}c@{}}RGB \\ images\end{tabular} & \begin{tabular}[c]{@{}c@{}}LiDAR \\ PCs\end{tabular} & \begin{tabular}[c]{@{}c@{}}Radar \\ frames\end{tabular} & Night/Rain & Locations               \\ \hline \hline
	CamVid \cite{Brostow_CamVid_2008}                & 2008          & \multirow{13}{*}{\begin{tabular}[c]{@{}c@{}}Autonomous \\ Vehicle\end{tabular}} & No                                                        & 4         & 0.4       & 18k                                                   & 0                                                    & 0                                                    & No/No      & Cambridge               \\ \cline{1-2} \cline{4-11} 
	KITTI\cite{Geiger_KITTI}                 & 2012          &                                                                                 & No                                                        & 22        & 1.5       & 15k                                                   & 15k                                                  & 0                                                    & No/No      & Karlsruhe               \\ \cline{1-2} \cline{4-11} 
	Cityscapes\cite{cordts2016cityscapes}           & 2016          &                                                                                 & No                                                        & n/a       &           & 25k                                                   & 0                                                    & 0                                                    & No/No      & $50 \times$ Germany               \\ \cline{1-2} \cline{4-11} 
	BDD100K\cite{yu2018bdd100k}               & 2017          &                                                                                 & No                                                        & 100k      & 1k        & 100M                                                  & 0                                                    & 0                                                    & Yes/Yes    & USA (NY, SF)                  \\ \cline{1-2} \cline{4-11} 
	ApolloScape\cite{huang2019apolloscape}           & 2018          &                                                                                 & No                                                        & -         & 100       & 144k                                                  & 0                                                    & 0                                                    & Yes/No     & $4 \times$ China        \\ \cline{1-2} \cline{4-11} 
	AS LiDAR\cite{ma2019trafficpredict}              & 2018          &                                                                                 & No                                                        & -         & 2         & 0                                                     & 20k                                                  & 0                                                    & -/-          & China                   \\ \cline{1-2} \cline{4-11} 
	H3D\cite{patil2019h3d}                   & 2019          &                                                                                 & No                                                        & 160       & 0.77      & 83k                                                   & 27k                                                  & 0                                                    & No/No      & USA (SF)                      \\ \cline{1-2} \cline{4-11} 
	nuScenes\cite{caesar2020nuscenes}              & 2019          &                                                                                 & No                                                        & 1k        & 5.5       & 1.4M                                                  & 400k                                                 & 1.3M                                                 & Yes/No     & $3 \times$ USA, SG      \\ \cline{1-2} \cline{4-11} 
	Argoverse\cite{chang2019argoverse}             & 2019          &                                                                                 & No                                                        & 113       & 0.6       & 490k                                                  & 44k                                                  & 0                                                    & Yes/Yes    & Miami, PT               \\ \cline{1-2} \cline{4-11} 
	Lyft L5\cite{houston2020thousand}               & 2019          &                                                                                 & No                                                        & 366       & 2.5       & 323k                                                  & 46k                                                  & 0                                                    & No/No      & Palo Alto               \\ \cline{1-2} \cline{4-11} 
	Waymo Open\cite{llc2019waymo}            & 2019          &                                                                                 & No                                                        & 1k        & 5.5       & 1M                                                    & 200k                                                 & 0                                                    & Yes/Yes    & $3 \times$ USA          \\ \cline{1-2} \cline{4-11} 
	A*3D \cite{pham20203d}                 & 2019          &                                                                                 & No                                                        & n/a       & 55        & 39k                                                   & 39k                                                  & 0                                                    & Yes/Yes    & SG                      \\ \cline{1-2} \cline{4-11} 
	CRUW \cite{wang2021rethinking}                 & 2021          &                                                                                 & No                                                        & -        & 3         & 396k                                                  & 0                                                    & 396k                                                 & -/-          & China                   \\ \hline \hline
	DAIR-V2X \cite{yu2022dair}             & 2022          & V2X                                                                             & No                                                        & -         & -         & 71k                                                   & 71k                                                  & -                                                    & -/-          & China                   \\ \hline
	\textbf{DeepSense 6G} & \textbf{2023} & \textbf{V2V}                                                                    & \textbf{Yes}                                              & \textbf{630} & \textbf{3.5} & \textbf{756k}                                             & \textbf{126k}                                            & \textbf{524k}                                            & \textbf{Yes/No}  & \textbf{Tempe, AZ, USA} \\ \hline
	\end{tabular}%
}
	\label{tab:literature_review}
\end{table*}

Motivated by the need for high-quality datasets specifically tailored for V2V communication research, we present the DeepSense 6G V2V dataset, the world's first large-scale real-world multi-modal sensing and communication dataset designed to facilitate V2V communication research and algorithm development. The DeepSense 6G V2V dataset is (i) a large-scale dataset of \textbf{more than 125k data points}, (ii) based on \textbf{real-world measurements}. The dataset comprises co-existing and synchronized \textbf{multi-modal sensing and communication data} and is organized in a \textbf{collection of 4 scenarios} captured from a diverse range of driving conditions and environments. These scenarios encompass urban, suburban, and rural highway settings, incorporating different traffic densities and road and weather conditions.

The DeepSense V2V dataset provides several key features that are essential for advancing V2V communication research:
\begin{itemize}
	\item \textbf{Co-existing sensing and communication:} The DeepSense V2V dataset consists of a large-scale collection of V2V mmWave communication data integrated with multi-modal sensing information. This unique combination empowers researchers to gain comprehensive insights into V2V scenarios, enabling them to explore the intricate interactions between sensor modalities and communication systems.
	
	\item \textbf{Co-located 360-degree sensor coverage:} The DeepSense V2V dataset leverages a diverse sensor suite, including cameras, radar, LiDAR, positioning sensors, and mmWave communication devices, to provide a $360$-degree coverage around the vehicle. This integration of different sensor modalities enables a comprehensive understanding of the surrounding environment, capturing rich data from visual observations, object detection, depth perception, positioning, and wireless communication dynamics. Moreover, the co-location of the sensors allows researchers to correlate sensory data better. 
	
	\item \textbf{Real World diverse scenarios:} The DeepSense V2V dataset is collected in real-world environments, providing a realistic representation of V2V communication scenarios in different locations, weather conditions, lighting settings, and traffic conditions. The dataset accurately captures real-world complexities and incorporates varying traffic densities, road conditions, and environmental influences.
	
	\item \textbf{Large-scale data:} Developing deep learning solutions that are scalable and robust to data distribution shifts (due to changes in the environment or deployment) requires the availability of a large-scale dataset. The DeepSense V2V dataset provides a large-scale collection of multi-modal data samples, comprising more than $125$k data points across four scenarios. This dataset's large-scale nature can help develop and evaluate advanced algorithms such as generalizability, robustness to distribution shift, etc.
\end{itemize}

This paper presents a detailed description of the DeepSense 6G V2V dataset, including its acquisition methodology, data formats, available scenarios, and annotations. Furthermore, we provide example use cases and highlight potential applications of the dataset in V2V communication research and algorithm development.

\section{Literature Review} \label{sec:Literature}

In recent years, publicly available datasets \cite{Brostow_CamVid_2008, Geiger_KITTI, cordts2016cityscapes, yu2018bdd100k, huang2019apolloscape, ma2019trafficpredict, patil2019h3d, chang2019argoverse, caesar2020nuscenes, llc2019waymo, pham20203d, wang2021rethinking, yu2022dair, choi2018kaist} have played a significant role in advancing the development of autonomous vehicle technologies. A summary of some of these key datasets is provided in Table ~\ref{tab:literature_review}. These datasets typically include data from various sensors, such as cameras, LiDARs, and GPS/IMU. They are often used for tasks such as object detection and segmentation, scene understanding, and localization and mapping. The KITTI dataset \cite{Geiger_KITTI}, with over 22 scenes, has been widely used for testing machine learning algorithms for vision tasks, such as object detection, using LiDAR and camera data. It provides 2D and 3D annotation data and has about 80k 2D and 3D bounding boxes. The H3D dataset \cite{patil2019h3d} includes 160 crowded scenes with 27k frames, with objects annotated in the full 360 views. The KAIST multi-spectral dataset \cite{choi2018kaist} is a multi-modal dataset comprising RGB and thermal cameras, RGB stereo, 3D LiDAR, and GPS/IMU, providing nighttime data. However, its size is limited. The NuScenes dataset \cite{caesar2020nuscenes} contains 1.4 million images and 400k point clouds collected from a sensor suite, including six cameras, one LiDAR, and five radars. It has 3D bounding box annotation, and its perception system mainly relies on LiDAR rather than cameras. The Waymo dataset \cite{llc2019waymo}, one of the largest and most diverse multi-modal autonomous driving datasets, contains 12 million 3D bounding boxes and 9.9 million 2D bounding boxes from its 1150 scenes, captured using 5 high-resolution cameras and 5 high-quality LiDARs. Its detection and tracking mainly rely on LiDAR rather than cameras, but the field-of-view (FoV) of the camera is less than 270º. A more detailed comparison of these datasets can be found in Table~\ref{tab:literature_review}.

These datasets can be used to evaluate and compare the performance of different algorithms and systems, which is important for advancing the state-of-the-art in autonomous vehicle technologies. The availability of large-scale datasets, especially in machine vision, allowed researchers to design more accurate and robust approaches and get a step closer to full driving autonomy. However, the existing datasets predominantly consist of a single vehicle collecting all the data and are unsuitable for vehicle-to-vehicle (V2V) collaborative applications. Collaboration between vehicles has been envisioned to play an important role in the personal mobility paradigm. For example, V2V communications enable collision warnings \cite{collision_warning_v2v}, which can prevent $60\%$ of road accidents according to some studies \cite{collision_avoidance_60_percent}. Another example is peer-to-peer data sharing, particularly streamed video, aimed at reducing the load in the wireless infrastructure when all vehicles require the same data \cite{p2p_video_streaming}, a common scenario in broadcasting events like football/soccer games.

To answer the need for V2V-specific real-world data, we introduce DeepSense-V2V, the first large-scale dataset for sensing, localization, and communications in V2V communication scenarios. It is a multi-modal dataset comprising data from mmWave wireless communication, GPS, vision, Radar, and LiDAR, all collected in a real-world wireless environment. In the following section, we present the DeepSense V2V dataset in detail. 

\begin{figure}[t]
	\centering
	\includegraphics[width=1\columnwidth]{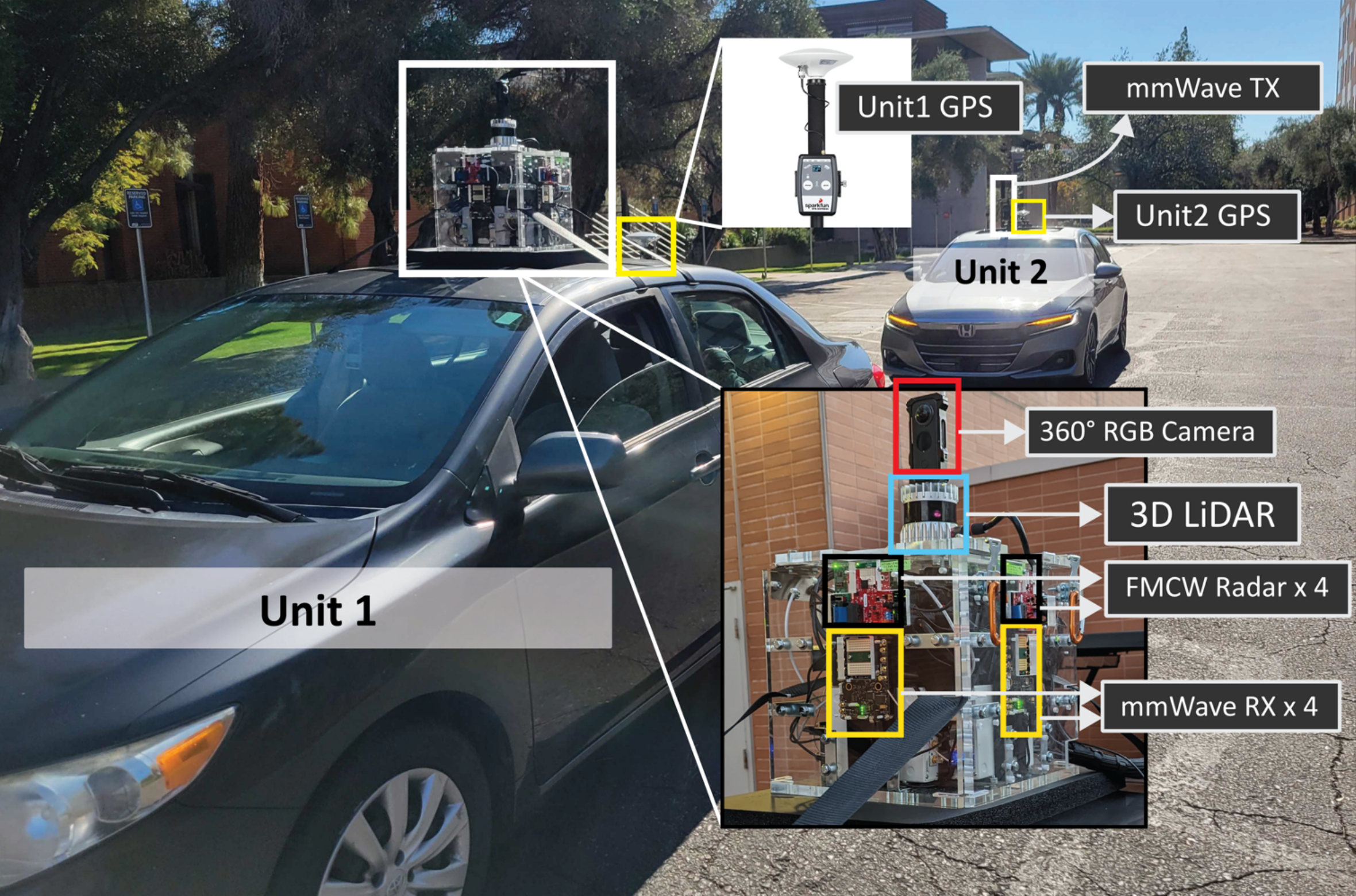}
	\caption{DeepSense V2V testbed setup overview. For more information on the testbed visit: \href{https://www.deepsense6g.net/data-collection/}{Testbed6}}
	\label{fig:V2V_testbed}
\end{figure}

\section{DeepSense V2V Testbed and Scenario Creation}

The V2V scenarios in DeepSense6G \cite{DeepSense} leverage a two-vehicle testbed. Car/unit 1 is the receiver and is equipped with four mmWave phased arrays facing four different directions, a 360-degree RGB camera, four mmWave FMCW radars, one 3D LiDAR, and one GPS RTK kit. Car/unit 2 is the transmitter and is equipped with a mmWave quasi-omnidirectional antenna always oriented towards the receiver and a GPS receiver to capture real-time position information. Figure \ref{fig:V2V_testbed} illustrates the composition of the testbed. This section describes the steps to acquire data from the sensors and process the data into this dataset. In particular, the data capture/sampling is detailed in Section \ref{subsec:Collection}. The key processing steps are described in \ref{subsec:Processing}. The processing procedure is verified via synchronized visualizations of all data, addressed in section \ref{subsec:Visualization}. Next, we detail the structure of how these phases of scenario creation come together, as well as their vital components. 

\begin{figure}[t]
	\centering
	\includegraphics[width=1\columnwidth]{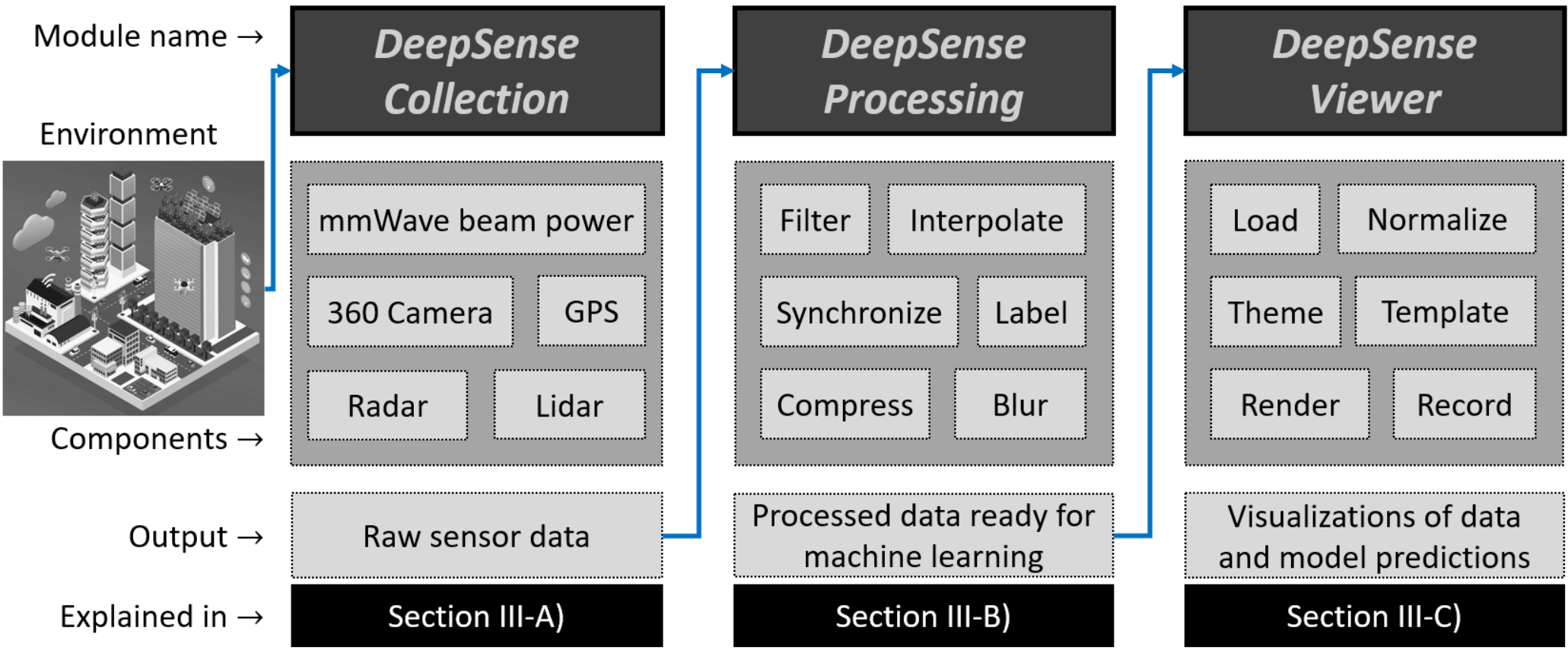}
	\caption{Overview of general DeepSense structure that was used in the creation of the V2V Scenarios.}
	\label{fig:deepsense_overview}
\end{figure}

\textbf{DeepSense Structure:} DeepSense scenario creation follows a general structure illustrated in the figure \ref{fig:deepsense_overview}. A general structure allows full automation of most tasks in the scenario creation pipeline, which in turn leads to (a) higher data quality: less prone to human error; (b) more reproducibility: the processing method is accurately coded; and (c) better scalability: since the process is automated, tasks are easier to execute, and the cost of adopting more challenging use-cases is reduced. These advantages become crucial requirements when data collection efforts grow to the size of the V2V scenarios presented in this paper. The structure comprises three stages coded into three large Python libraries: DeepSense Collection, DeepSense Processing, and DeepSense Visualization. These libraries build on top of popular high-performance scientific computing tools. A short description of the three stages follows:
\begin{itemize}
    \item \textbf{DeepSense Collection}: Responsible for transducing environment information into sensor data. 
    \item \textbf{DeepSense Processing}: Responsible for converting, filtering, interpolating, and synchronizing the raw sensor data into a processed DeepSense scenario.
    \item \textbf{DeepSense Visualization}: Used to aid and verify the processing stage and to render scenario videos. 
\end{itemize}
In the following subsections, we will break down the stages in order to clarify how the dataset was constructed.

\subsection{Data Collection} \label{subsec:Collection}

\begin{figure}[t]
	\centering
	\includegraphics[width=1\columnwidth]{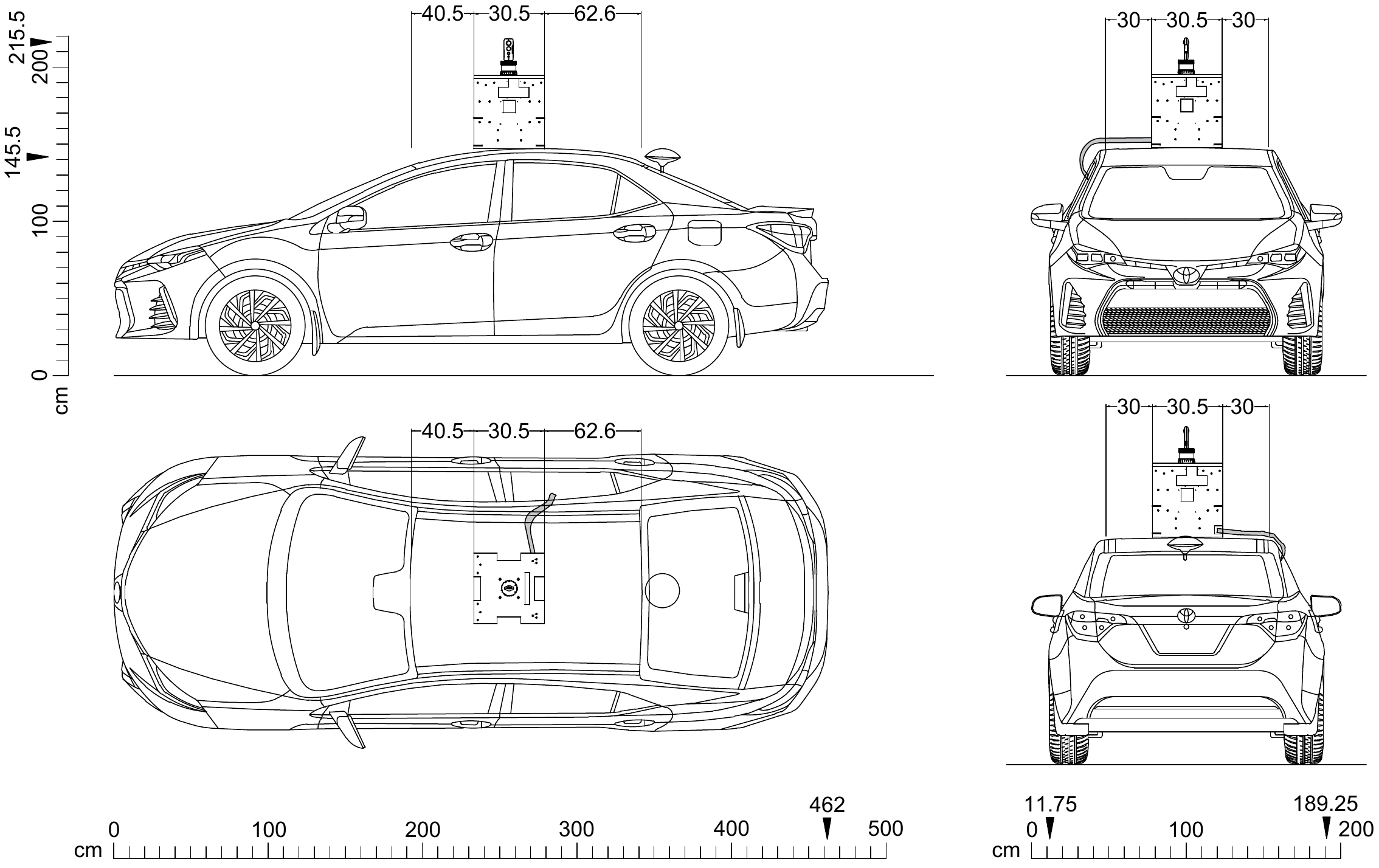}
	\caption{CAD design with dimensions of V2V box placement on car.}
	\label{fig:sensor_testbed_schematic}
\end{figure}

The data collection stage comprises all the software and parameter configurations needed to collect data from the sensors present in the two units. Car/unit 1 (the car in front in Figure \ref{fig:V2V_testbed}) contains the V2V box, a half-inch thick acrylic enclosure that holds all the sensors except the GPS. The sensors in the box are carefully detailed in this section, but the box fabrication procedure is omitted for brevity. Car/unit 2 (in the back in Figure \ref{fig:V2V_testbed}) consists of the same GPS fixed on the vehicle and a phased array mounted on a tripod. A schematic of the dimensions of the V2V box and its position in the car is shown in Figure \ref{fig:sensor_testbed_schematic}. This section describes the sensors that generated the data in this dataset and the collection context in which data was acquired. 

\textbf{Sensor Suite:} It comprises different sensors with different functions and limitations, as well as different sampling times and physical interface requirements (i.e., for power and connectivity). All non-communication sensors - the four radars, the 3D lidar, the 360º camera, and the two GPSs - operate in continuous data acquisition mode with a predefined sample rate. This is not the case with the mmWave beam power collection, where the receiver radio and phased arrays are programmatically triggered to collect a sample every 100 ms. A beam power sample consists of a sweep of the 64 beams spanning -45 to +45 degrees in azimuth and measuring the received power in each of those beams. This 64-valued power vector is our unitary sample for communications. 

Besides the mmWave beam powers, the testbed comprises 2 GPSs using the L1 and L2 bands for higher accuracy - the horizontal accuracies are always within a meter of the true, according to manufacturer information and horizontal dilution measures returned by the device. The testbed also holds a 360º video camera, which is used to export four 90º views and two 180º views around the car, effectively covering all angles and emulating the existence of multiple cameras around the vehicle. The single lidar in the testbed creates a 32 thousand-point 3D point cloud with a maximum range of 200 meters. In terms of range, the configurations of the four radars allow more than 200 maximum distance, but factors like clutter and ADC resolution prevent such ranges in realistic road situations. More information on the sensors, like each sample rate, the location of the sensors in each unit, specific resolutions, and configurations, can be consulted in Table \ref{tab:sensors}. 

\textbf{Collection Procedure:} The data was acquired in the following way. First, all the sensors are initialized at the start of collecting data. The mmWave power captured by the box in unit 1 comes from an omnidirectional transmitter in car/unit 2. This transmitter is attached to a tripod and is manually rotated to guarantee power at the receiver (unit 1). The system is capable of displaying the power received in each beam in real time. This monitoring capability is used mainly to start vehicle movement once a received power vector is visually verified. The trajectory is coarsely planned ahead of time. The two vehicles attempt to stay relatively close throughout the collection such that the received power in the optimum beam is higher than the noise floor. As the distance grows, the blockages also become more likely. Nonetheless, in LoS conditions, the received power in the best beam is distinguishable from noise over 500-meter distances. This distance is more likely achieved in V2I situations. For example, in a V2I situation, the box can play the role of a basestation or be placed in the car to communicate with a static unit that acts as the BS. Effectively, the testbed described here can be used in a range of V2X applications.

\begin{table*}[t]
\centering
\caption{Description of the sensors used in the DeepSense-V2V testbed.}
\label{tab:sensors}
\begin{tabular}{|c|c|c|c|l|}
\hline
Modality                                                         & Sensors                                                                                     & Quantity                                                    & Sample Rate & \multicolumn{1}{c|}{More sensor information and remarks}                                                                                                                                                                                                                                                                                                                          \\ \hline
\begin{tabular}[c]{@{}c@{}}mmWave \\ Beam \\ Powers\end{tabular} & \begin{tabular}[c]{@{}c@{}}Sivers Phased \\ Array \\ (EVK06003) \\ + USRP B210\end{tabular} & \begin{tabular}[c]{@{}c@{}}unit1: 4\\ unit2: 1\end{tabular} & 10 Hz                                                             & \begin{tabular}[c]{@{}l@{}}- Unit 1: receive mode, sweeping codebook of 64 beams. \\ - Unit 2: transmit mode, near-omnidirectional \\ - Phased arrays: 16-element ULA with 62.64 GHz center frequency\\ - Phased arrays: up/downconvert zero IF to/from the USRP \\ - USRP: 640 samples per beam at 5 MHz sample rate \end{tabular} \\ \hline
GPS                                                              & RTK Express                                                                                 & \begin{tabular}[c]{@{}c@{}}unit1: 1\\ unit2: 1\end{tabular} & 10 Hz                                                             & \begin{tabular}[c]{@{}l@{}}- Accuracy within 0.5m (\textgreater{}90\% of the time)\\ - Easy to interpolate\end{tabular}                                                                                                                                                                                                                                                          \\ \hline
Image                                                            & \begin{tabular}[c]{@{}c@{}}360º Camera\\ (Insta 360 One X2)\end{tabular}                    & \begin{tabular}[c]{@{}c@{}}unit1: 1\\ unit2: 0\end{tabular} & 30 Hz                                                             & \begin{tabular}[c]{@{}l@{}}- Sensitive to lighting conditions\\ - Individual images (90º and 180º views) are rendered \\ from a 360º video\\ - 5.7 K resolution\end{tabular}                                                                                                                                                                                                     \\ \hline
Radar                                                            & \begin{tabular}[c]{@{}c@{}}AWR2243BOOST \\ + DCA1000EVM\end{tabular}                        & \begin{tabular}[c]{@{}c@{}}unit1: 4\\ unit2: 0\end{tabular} & 10 Hz                                                             & \begin{tabular}[c]{@{}l@{}}- Radar configurations: 128 chirps, 1 tx antenna, 4 rx, \\ 256 samples per chirp, 2 bytes per sample, 5 MHz ADC \\ sample rate, 15.015 THz/s chirp slope, 77 GHz frequency, \\ 60 us ADC start time, 5 us idle time\end{tabular}                                                                                                                      \\ \hline
Lidar                                                            & \begin{tabular}[c]{@{}c@{}}Ouster OS1 \\ 32 beams\end{tabular}                              & \begin{tabular}[c]{@{}c@{}}unit1: 1\\ unit2: 0\end{tabular} & 20 Hz                                                              & \begin{tabular}[c]{@{}l@{}}- 1024 horizontal beams (across 360º)\\ - 32 vertical beams (-45, +45º)\end{tabular}                                                                                                                                                                                                                                                                   \\ \hline
\end{tabular}
\end{table*}

\subsection{Data Processing} \label{subsec:Processing}

The intermediate stage of DeepSense scenario creation is data processing. While DeepSense Collection deals with data acquisition from sensors, often involving manufacturer-specific caveats, DeepSense Processing deals more generally with processing data formats independently of the sensor they come from. The data processing stage consists of two major phases:
\begin{itemize}
    \item Phase 1: converts data from the sensors of all modalities in timestamped samples. For example, a data capture with the lidar sensor is usually saved in a single file unsuitable for proper data synchronization. This phase takes care of extracting all samples and metadata for the sensor-specific data format and organizes them in clear CSVs. It may further interpolate data points (currently only in GPS).
    \item Phase 2: filters, organizes, creates sequences of continuous data acquisition, and synchronizes the extracted data into a processed DeepSense scenario.
\end{itemize}
Phase 1 processes different modalities in parallel, with specific steps tailored to each modality. For instance, GPS samples in the NMEA protocol format require different processing than video data from a 360º camera. While detailed descriptions of Phase 1 are beyond the scope of this discussion, it is essential to note that data and metadata are extracted from their original formats into a common structure suitable for ingestion and synchronization in Phase 2. Phase 2, unlike Phase 1, processes data sequentially and is agnostic to data formats. This phase focuses on data synchronization, filtering, sequencing, labeling, and compression. This discussion will primarily concentrate on the functions of Phase 2.

\textbf{Synchronization:} The synchronization step takes sensor data sampled at different time instants and different sample rates and obtains a uniform set of samples at a single sample rate. At its core, the synchronization process is a one-to-one sample mapping based on timestamp proximity. In more detail, the first step is selecting the right sample rate. The sample rate used in the V2V scenarios is 10 Hz. The next step is choosing a reference modality to dictate the sampling intervals the other sensors should attempt to approximate. This reference modality is the mmWave power. Then, for each sampling interval, the synchronization stage chooses the closest sample of each modality to this instant. All the samples not selected for any sampling instant will be discarded. For example, RGB images are sampled at 30 Hz but Power only at 10 Hz; roughly two-thirds of images will be discarded in this step. 

\textbf{Filtering} involves rejecting samples according to a set of criteria. It happens during synchronization due to oversampling, and it happens in three other situations: a) due to acquisition errors, like blank or repeated samples; b) due to non-coexistence, i.e., when sensors are not sampling at the same time due to problems or human errors during the collection, or c) sequence filtering, as we describe next. 

\textbf{Sequencing} is the task of separating samples into groups of continuous samples. Samples in the same sequence tell the user that those samples were acquired precisely 0.1 seconds apart. This is necessary because sensor failures, human error, and other problems can lead to a continuity break, resulting in gaps larger than 0.1 seconds between samples. When sampling continuity is broken, the sequence ends and a new one starts when continuity is achieved again. It is relevant to mark sample continuity in the dataset because several downstream (ML) tasks depend on this continuity. DeepSense accurately records continuity disruptions to be effectively used in these tasks.

\textbf{Data labels} are additional information that can be useful for the use of the dataset. These labels can be extracted from the sensor. For example, GPS reports the number of satellites and position errors, and this information can be useful for research on localization. Labels can also be derived from sensor data; e.g., the best beam label is derived by computing the index of the beam with higher power. Or labels may even be manually added, as is the case with ground truth blockage labels or ground truth bounding boxes where the label is directly added by humans or obtained from processes with humans in the loop. Overall, labels provide extra contextual information useful in certain researcher use cases.

\textbf{Data compression} is performed for more efficient, flexible, and robust distribution. Data is compressed in 8 GB parts using the 7zip utility with the level 5 deflate method. The result is a significant reduction in the number of files and the total size, which consequently leads to users of the dataset being able to download the dataset faster and more reliably. The compression stage also separates into different files the different modalities. Therefore, researchers may download only the modalities of interest. 

\begin{figure*}[t]
	\centering
	\includegraphics[width=2\columnwidth]{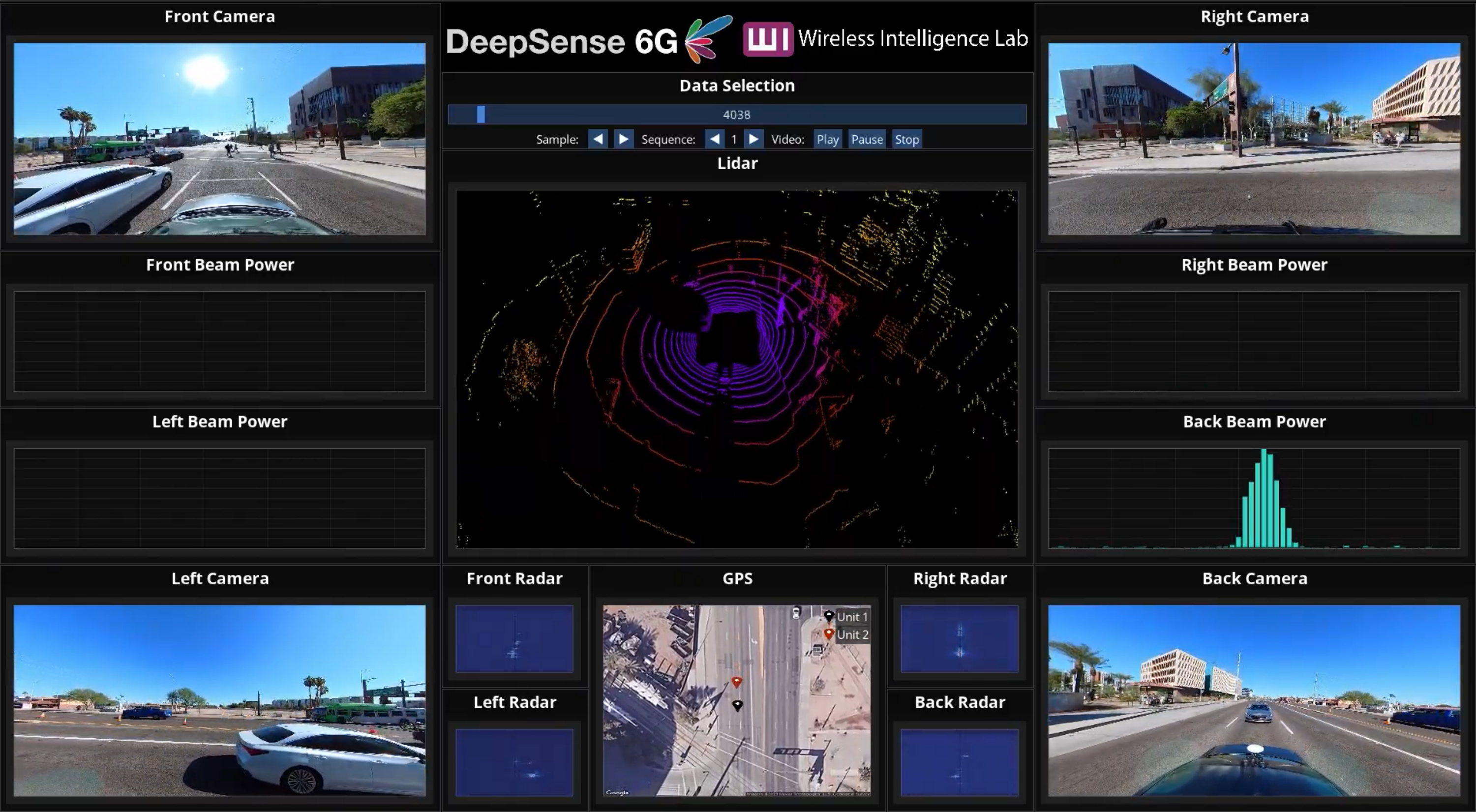}
	\caption{Frame of sample 4038 from video of Scenario 36. The current template shows four 90º camera views rendered around the car, the lidar pointcloud colored based on distance, four radar range-velocity plots, a GPS with the locations of the vehicles scattered on top of the satellite image of the location, and four 64-beam power vectors with the normalized received power in each beam. The video rendered for Scenario 36 data can be watched on \href{https://www.youtube.com/watch?v=9RyZnZI7kv0&ab_channel=DeepSense6G}{YouTube}}
	\label{fig:data_visualization}
\end{figure*}

\textbf{Other data modifications} refer to adjustments that do not fall within the previously defined categories, and currently, there are only two such modifications. The first is interpolation, the insertion of generated data derived from true data points before and after the insertion. We interpolate to obtain data at the sampling intervals of the mmWave powers. Currently, we only interpolate GPS data. The GPS interpolation is linear and is clearly marked in the CSV file that indexes all data. The CSV normally contains labels with at most four decimal places, but the interpolated values will have 8. Linear position and GPS label interpolation are only conducted for distances less than 1 second apart. Less than 5\% of the GPS data across all scenarios is interpolated. Given the considered mobility profiles, we verified that interpolating intervals of 1 second still provide a very good approximation of reality. The second case where data modifications take place is to protect privacy. Although local law does not mandate face blurring in videos recorded in public places, we still do it for extra safety and to guarantee the wide usability of the dataset. Besides these two cases, no other data alteration steps are performed during data processing. This includes normalization, meaning that magnitudes in the dataset are preserved from the sensor. Although the dataset includes all original values, the data may be normalized for visualization. Next, we present the final stage, data visualization.

\subsection{Data Visualization} \label{subsec:Visualization}

Data Visualization provides significant value in several fronts: dataset interpretability and understanding, fast identification of the samples of interest, easier recognition of propagation phenomena, like reflections, blockages, large distances radio transmission, and easier spotting of adverse sensor conditions, such as hard visibility from light or weather and excessive radar clutter. To enable these advantages, the DeepSense scenario creation pipeline leverages a data visualization user interface (UI) in the DeepSense Viewer library. We use this UI to verify the individual stages of data processing and to render a final scenario video that synchronizes all processed data. An example of a scenario video is depicted in Figure \ref{fig:data_visualization}. This figure shows all modalities present in the dataset, including both units. Some modalities are normalized only to facilitate the visualization, namely by assuring relevant features are not hidden by ill-defined scales or less-clear colormaps. The data displayed in each frame of the video is from the same time instant and corresponds to one row of the indexing CSV. 

\textbf{Scenario videos:} Using the user interface built within the DeepSense Viewer module, we render a video for each scenario where data is displayed across time. In our experience, this video makes data easy to navigate and allows the researcher to find the moments of interest. These videos are rendered at four times the real-world speed to allow the user to visualize large portions of the dataset quickly. YouTube allows a 0.25x speed control that will bring the speed back to the real world, and for finer controls, the user can use keyboard shortcuts to navigate the video frame by frame - for this reason, the video is rendered to have a different sample in each frame. These videos can be found on the web page of the V2V DeepSense Scenarios (i.e., \href{https://deepsense6g.net/scenarios36-39/}{Scenarios36-39}).In this paper, however, we will show the variability and reach of the proposed dataset differently from videos. In the following section, we show interesting patterns and statistics that researchers can exploit for developing machine learning algorithms for V2V communications.

\section{Dataset Statistics} \label{sec:Statistics}

A useful dataset with wide applicability in wireless communications should contain substantial variability while being accurate and consistent. This section shows many statistics about the location and speed of the vehicles during the data collection in Section \ref{subsec:loc_and_vel}. Then it delves into how received power relates to distance in Section \ref{subsec:distance_vs_receivepower} to prove the consistency of data. Subsequently, mmWave and GPS data are again related when we display beam distributions and position distribution across time in \ref{subsec:beam_distributions}, showing that the direction of the incoming signal strongly correlates with the beam. This should be because LoS is the predominant link status during collection. Then, Section \ref{subsec:img_detection} shows the results obtained from applying machine vision detection and classification approaches to the visual data. This section illustrates the visual diversity in the dataset by showing a high volume of road-related objects identifiable throughout the dataset.

\subsection{Vehicle Locations and Velocities} \label{subsec:loc_and_vel}

\begin{figure}[t]
	\centering
    \includegraphics[width=1\columnwidth]{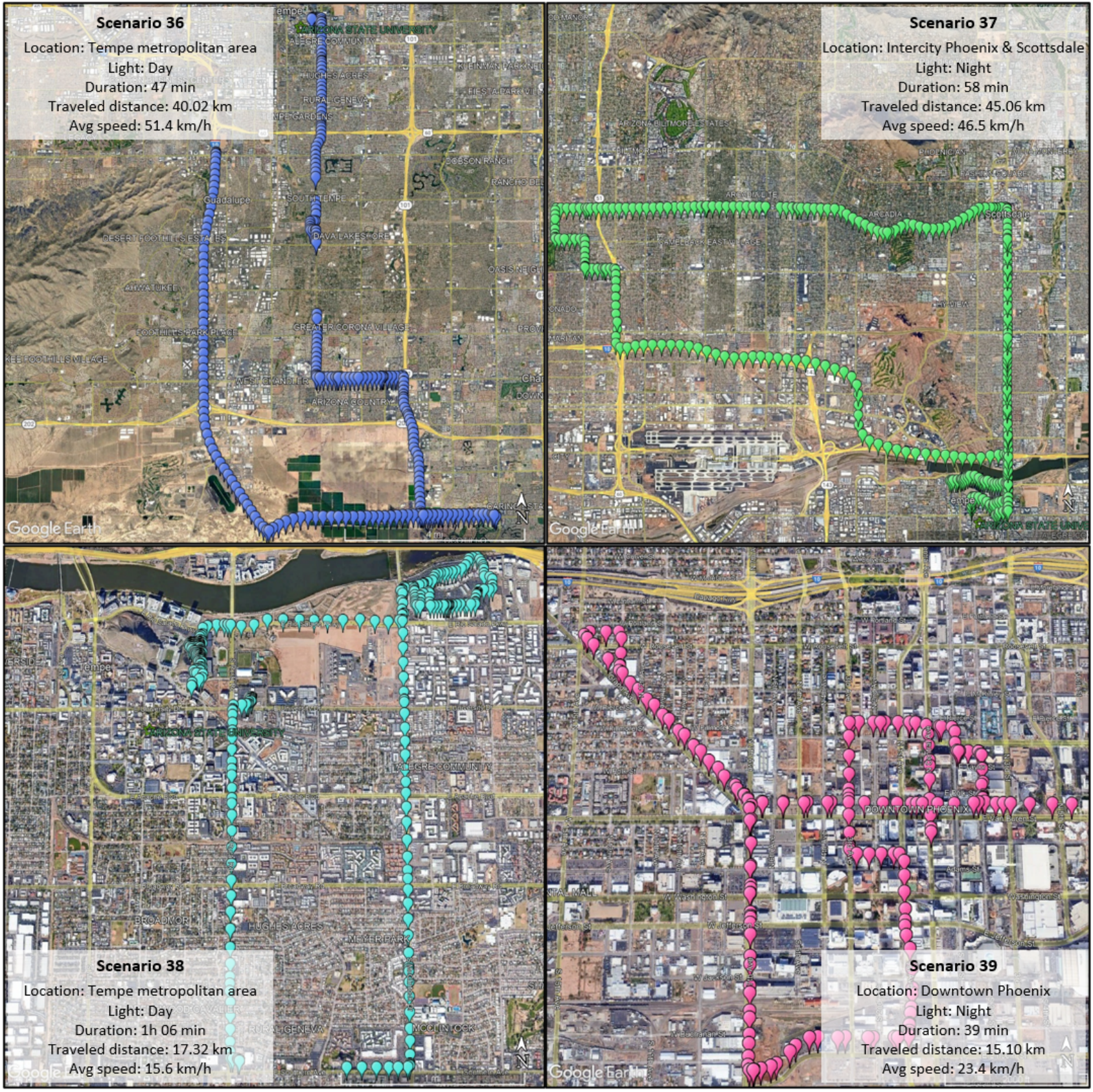}
	\caption{Satellite images with the locations of each scenario. Also included are several macro statistics of the data collection, providing contextual and objective information derived mainly from the GPS sensors.}
	\label{fig:locs}
\end{figure}

\begin{figure}[t]
	\centering
	\includegraphics[width=1\columnwidth]{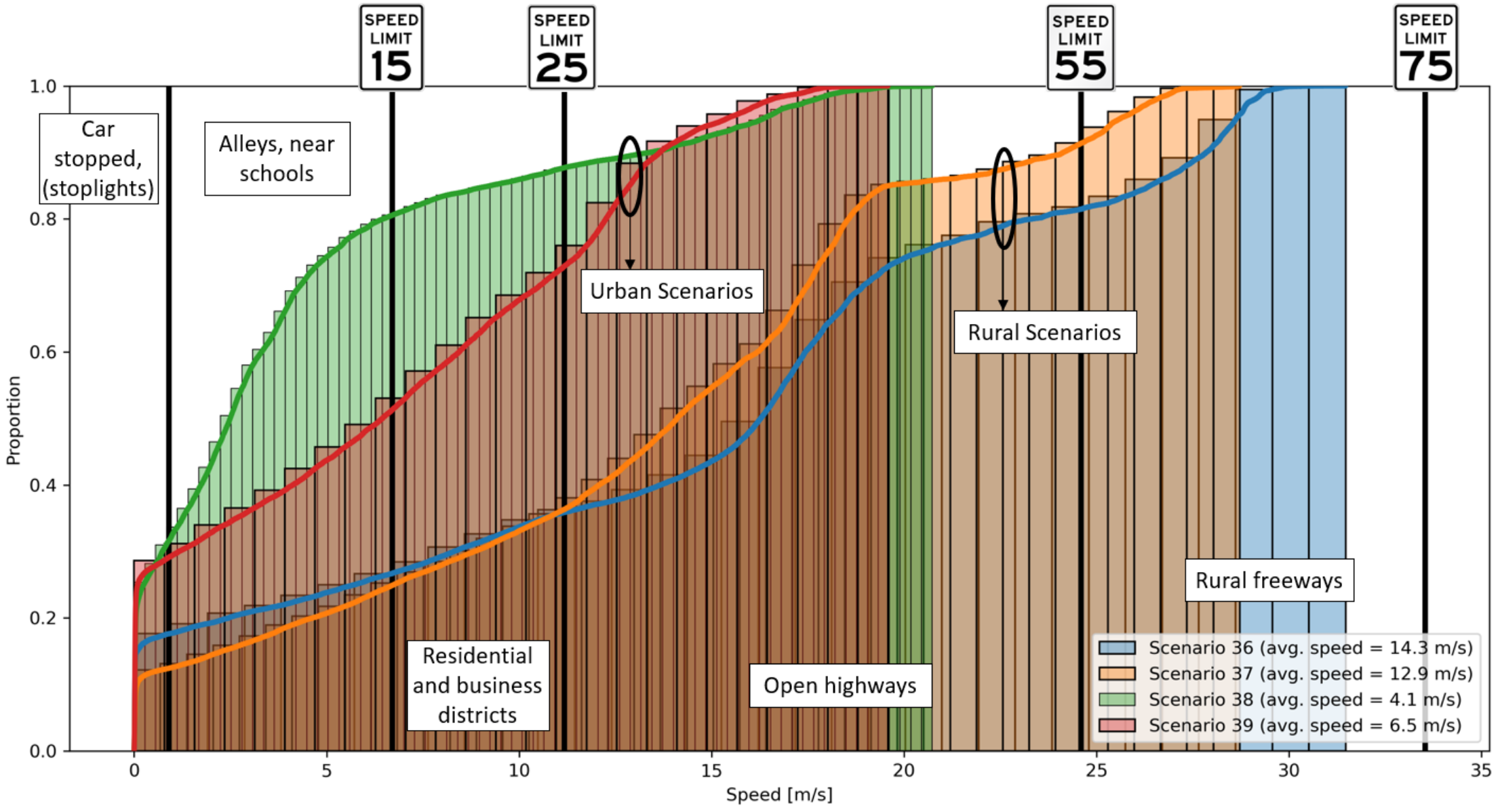}
	\caption{Speed cumulative distribution of vehicle 1 with the indication of the speed limits (in mph) and the type of road that matches the interval of speeds. }
	\label{fig:speeds}
\end{figure}

Vehicle locations play an essential role in the surroundings, which heavily impact propagation, thus affecting not only wireless communications but also GPS, Lidar, and Radar. In Figure \ref{fig:locs}, we illustrate the locations of the receiver captured by the GPS (undersampled by a factor of 100 to facilitate readability), along with other macro statistics of the data collection. Scenarios 36 and 37 are collected in long drives between cities, targeting long travels, while Scenarios 38 and 39 are more oriented to emulate short urban commutes, so data is predominantly inside cities. For this reason, we call Scenarios 36 and 37 inter-city scenarios and 38 and 39 urban scenarios. The difference is corroborated by the traveled distance and average speed. While Scenarios 36 and 37 have long-distance travel at relatively high average speeds, 38 and 39 traveled less at a lower speed because of speed limits within cities. We further look into speed distributions in Figure \ref{fig:speeds}.

Furthermore, we include information like the lighting and weather conditions relevant to accessing the capabilities of cameras versus lidars and radars. For completeness' sake, the time of the first and last samples are included to describe the span of the collection. Note, however, that during the data collection, there are intermittent pauses in the acquisition of data. This justifies why the span of the collection and the filtered duration of the collection often have different numbers, with the former bigger than the latter. Some reasons for such pauses can be associated with hardware limitations, like the need to change batteries in the 360 camera, or they can be associated with errors in the collection where cars got too far apart and the signal got interrupted for a long time, or when one of the sensors had an error and did not acquire data for some time. We opt not to include samples where all modalities are not present. 

\textbf{Speed} distributions can tell the diversity of vehicle movement speeds in the dataset. Moreover, given that the speed limits were closely followed during data collection, we can further extrapolate what kind of roads the vehicles were on from their speed. Information on the type of roads is relevant because it tells what kind of objects and phenomena we expect to find in those samples. Figure \ref{fig:speeds} shows each scenario's cumulative distributions of speeds. We can observe that the intercity / rural scenarios (36 and 37) have a more flat distribution with contributions from higher speeds than the urban scenarios (38 and 39). Higher speeds come from driving in free-ways, and very low speeds result from traffic lights, intersections, and stop signs, characteristics of dense urban mobility. We also indicate the speed limit regulations in Arizona, USA, in miles per hour. This information allows us to estimate, for example, that the car in Scenario 38 was stopped in traffic lights for over 20\% of the time and that the car in Scenario 39 was driven in alleys or in residential/business districts for about 50\% of the time.

\begin{figure*}[t]
	\centering
	\includegraphics[width=1.99\columnwidth]{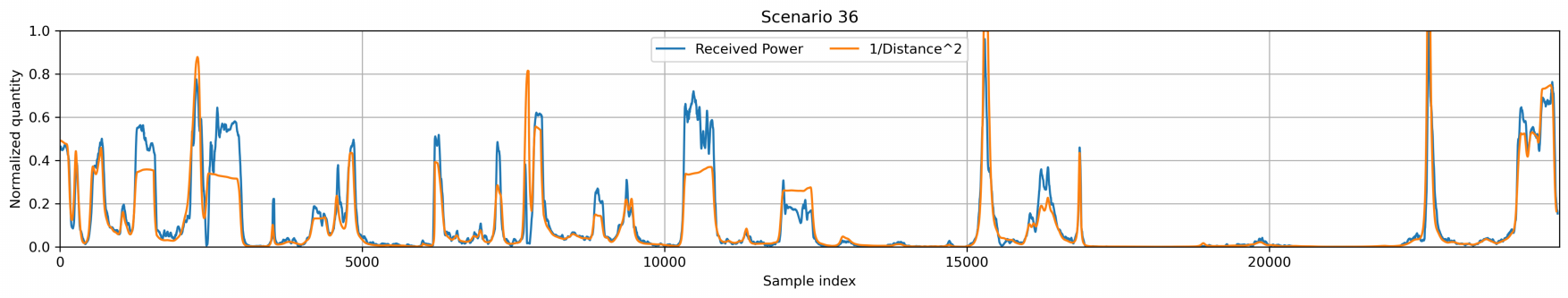}
	\caption{Relation of received power (blue) and the inverse of the distance between two vehicles square (in orange). The figure illustrates the relation between the two quantities, which always vary together when there is a LoS link between the two vehicles.}
	\label{fig:distance_vs_power}
\end{figure*}

\subsection{Inter-vehicle Distance and Received Power} \label{subsec:distance_vs_receivepower}

The distance between the receiver and transmitter and the received power in the optimum beam are closely related to the radio propagation theory of a LoS link. Since this dataset uses mmWave frequencies, which require a LoS in most cases, this dataset should reflect the power-distance relation. We show this relation across all scenarios by charting in Figure \ref{fig:distance_vs_power} the distance (or, more accurately, the inverse of the distance square) and the received power. The figure shows a strong correlation between distance and received power. But there also are cases where the correlation is broken (e.g., from sample 7500 to 8100 of Scenario 36) due to blockage and NLoS. Furthermore, it should be noted that the powers present in this dataset are not in Watts. We acquire baseband powers by computing the square of the amplitude of the baseband samples. Accurately measuring received powers at the antenna requires a difficult calibration process with both the receiver and transmitter. Instead, we attempted to perform data collection always within the linear regions of all components. As such, the relation between distance and received power should hold. This is suggested by the results in Figure \ref{fig:distance_vs_power}.

\subsection{Beam Distributions and GPS Positions} \label{subsec:beam_distributions}

\begin{figure*}[t]
	\centering
	\includegraphics[width=2\columnwidth]{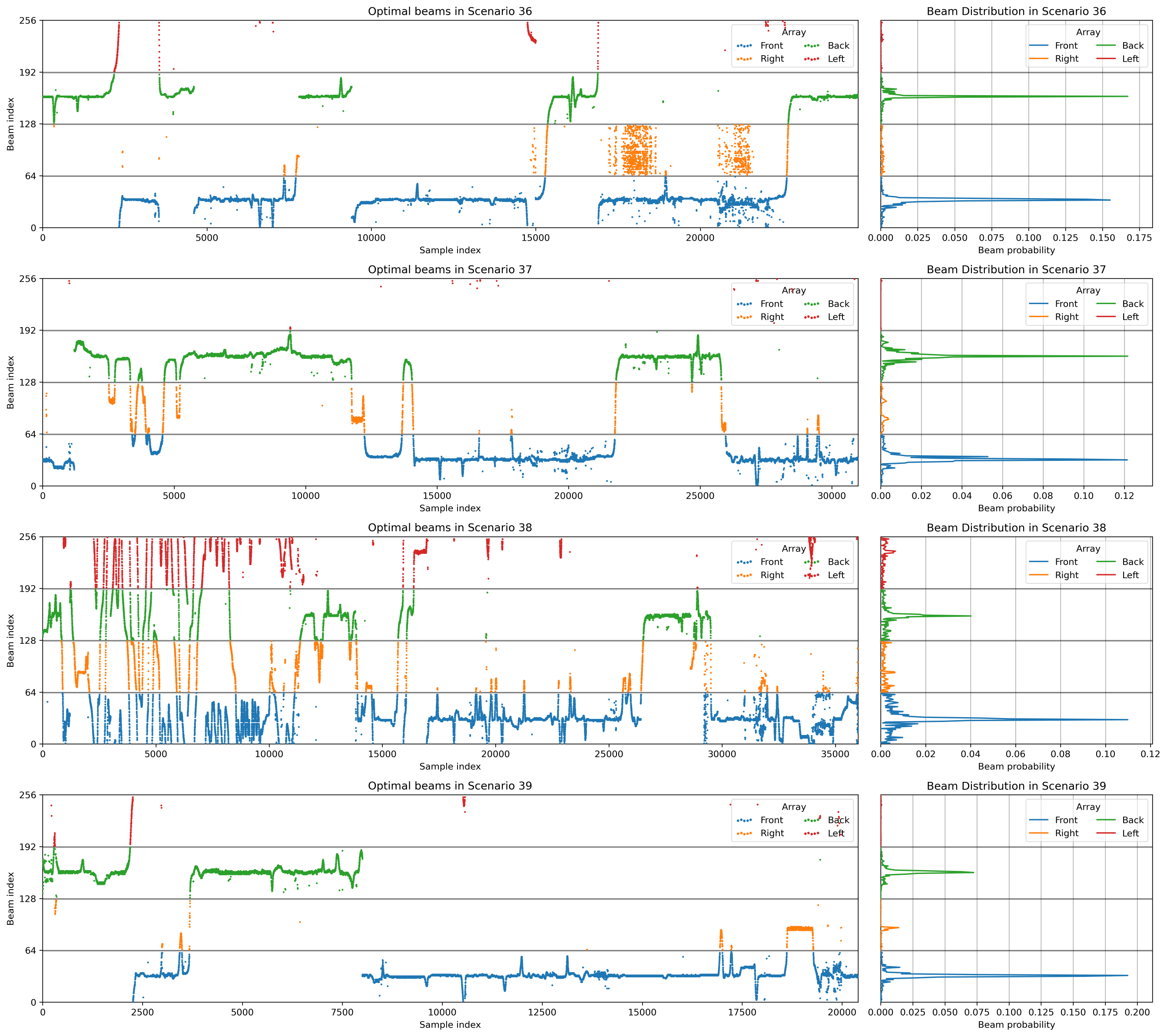}
	\caption{Optimal beam across time and corresponding beam distribution for all Scenarios show a tendency of vehicles driving in front or behind each other. Different colors represent beam indices on different phased arrays to provide panel-switching context information. Interruptions are due to sequence changing (see Section \ref{subsec:Processing}) or blockage due to other vehicles. In low SNR regimes, e.g., near sample 20000 of Scenario 36, the optimal beam becomes ambiguous.}
	\label{fig:beam_distro}
\end{figure*}

One differentiation factor of this dataset is that it includes beam information. Accordingly, we include Figure \ref{fig:beam_distro} that shows variations in the optimal beam across time and how they contribute to the overall beam density distribution. The figure also shows interesting phenomena. For example, given that most propagation in mmWave communications happens in LoS, we observe a continuous transition between beam indices. If beam continuity is interrupted, it can only be because of two reasons: i) the data collection was interrupted and the cars restarted in different positions, in which case we indicate that by changing sequences, since each sequence marks a continuous collection; ii) the second reason is when the cars get sufficiently far way NLoS or blockage. 

\begin{figure}[!t]
	\centering
	\includegraphics[width=1\columnwidth]{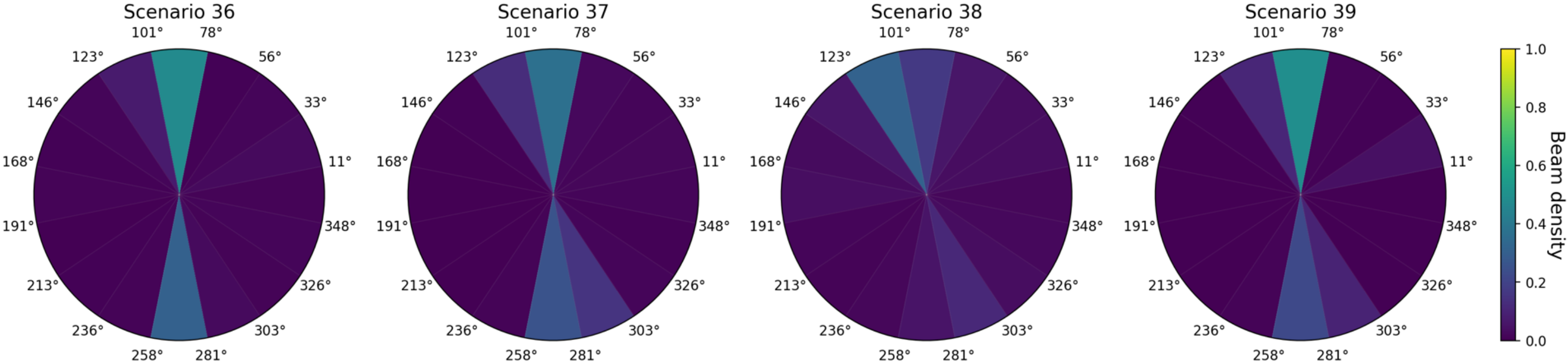}
	\caption{Beam density across angular space for all Scenarios.}
	\label{fig:beam_density}
\end{figure}

\begin{figure}[!t]
	\centering
	\includegraphics[width=1\columnwidth]{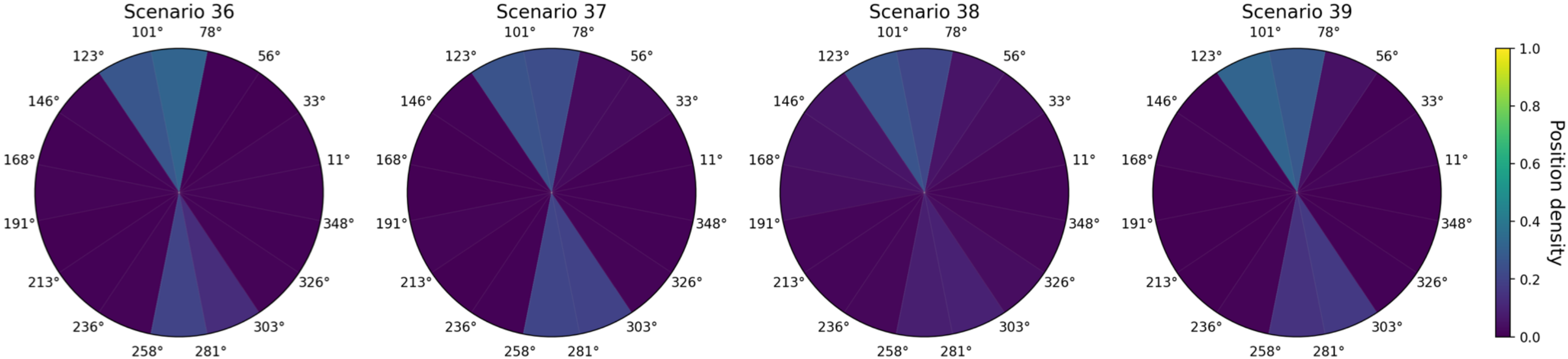}
	\caption{Relative orientation across angular space for all Scenarios.}
	\label{fig:pos_density}
\end{figure}

The visualization in Figure \ref{fig:beam_distro} also allows us to identify particular phenomena we might be interested in studying. Moreover, we color the beams from each panel with different colors; therefore, when we see that when a color changes (between indices 63/64, 127/128, and 191/192), it means a different panel or array is selected at the device (car). Also, in all scenarios, the beam distribution is concentrated in the middle of the front and back arrays. This is intuitive because two cars rarely spend long periods of time at the side of each other, but rather long times in front or at the back of each other. This is why there are long periods where the optimal beam lies in the middle of the front and back arrays (respectively colored in blue and green). We can also spot overtakes when we see a transition between front and back arrays, passing through the side arrays (colored in orange and red).

\textbf{Beam and Relative Position Densities:} It is essential to highlight the relation between beams and positions. We already showed this relation by relating the distance between the vehicles and the received power in Figure \ref{fig:distance_vs_power}. Now we highlight with respect to angle. Figures \ref{fig:beam_density} and \ref{fig:pos_density} show the distributions of beams in angle and relative positions between the two cars. Although it is not perfect, we see a strong correlation between the two. The relation is not perfect because of NLoS events and because the relative position is not always equivalent to the variable that should correlate perfectly with the optimal beam direction, the angle of arrival (AoA). Those situations happen when the receiver vehicle has a different orientation than the transmitter vehicle, thus changing the arrival angle without changing the relative position computed via GPS positions. In Section \ref{sec:ML_tasks}, we further augment our estimation of AoA to relate with beam choice more accurately. In the figure, we also see that the predominant beam directions and relative positions agree with the tendency for vehicles to drive in front or behind each other. 

\subsection{Machine Vision and Image Detection} \label{subsec:img_detection}

\begin{figure}[t]
	\centering
	\includegraphics[width=1\columnwidth]{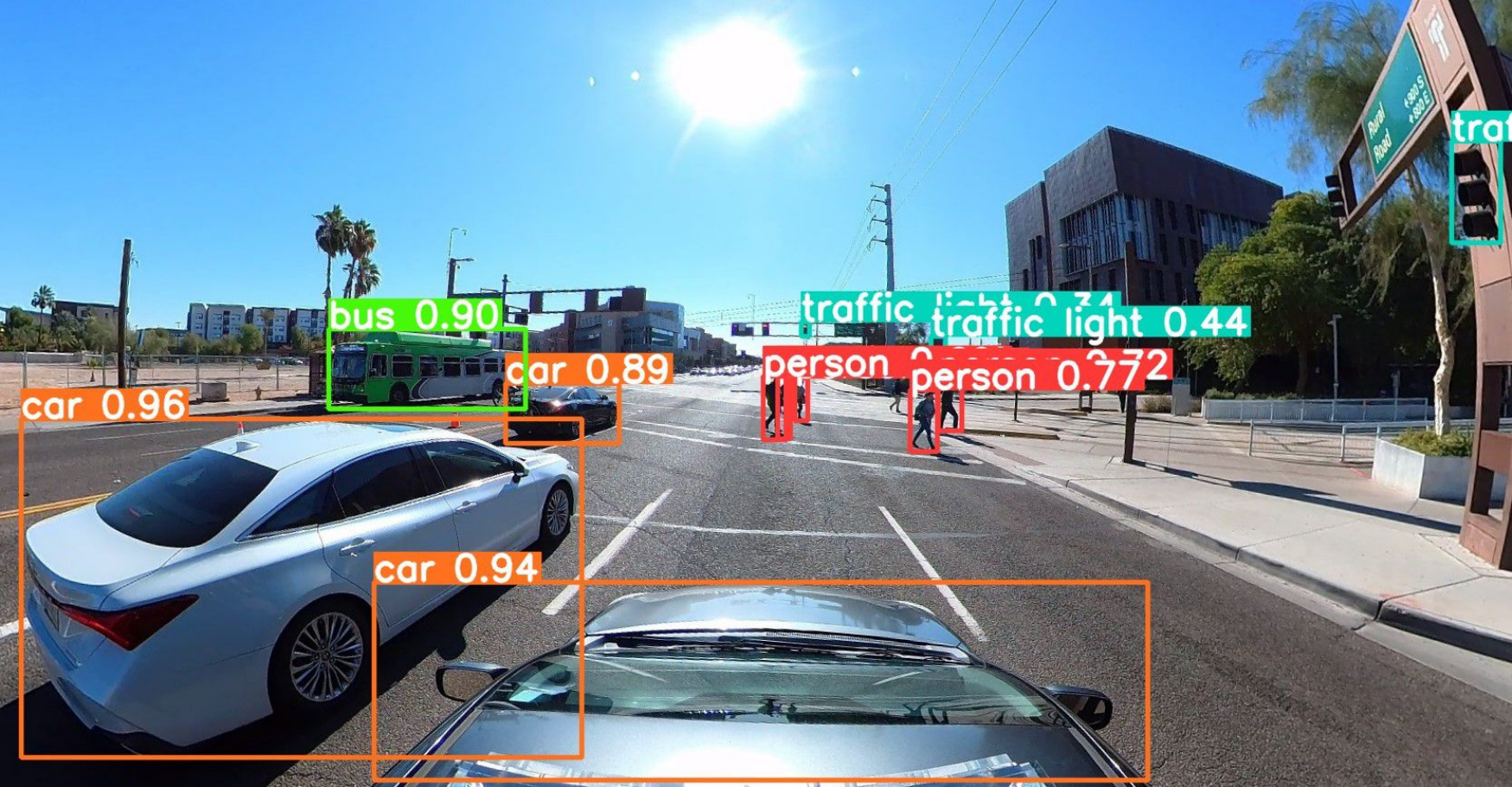}
	\caption{Example output from running YOLOv8 in image detection mode in a 180º front view image, belonging to sample 4035 of Scenario 36. The detection result is 5 people, 3 traffic lights, 2 cars (excluding ours), and a bus.}
	\label{fig:yolo_example}
\end{figure}

\begin{figure}[t]
	\centering
	\includegraphics[width=1\columnwidth]{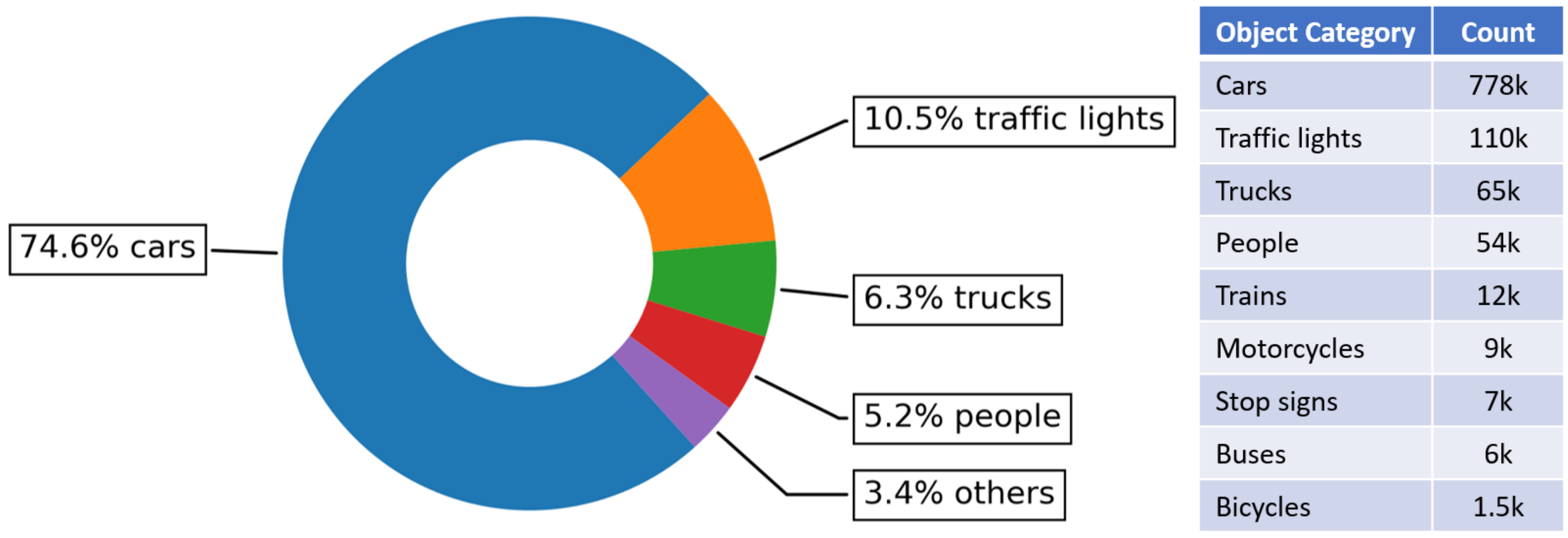}
	\caption{Results from running YOLOv8 in image detection mode in 250 thousand 180º images across all scenarios. On the left, a circular chart shows the classification percentage of the major categories. The table on the right presents finer detail in classification categories with the number of detections.}
	\label{fig:yolo_results}
\end{figure}

Modern cars, especially autonomous and semi-autonomous, already have cameras for several driving-related functions. To aid communications, for autonomous driving purposes, simply for security reasons or to increase the understanding of the environment, the content captured by cameras can be very useful. As such, we present in Figure \ref{fig:yolo_example} what a pre-trained state-of-the-art image model, YOLOv8 \cite{YOLOv8}, detects when enabled in detection mode. We executed the model to detect and classify objects in all 180º images of the dataset. These images were rendered from the 360º camera depicting the front and the back of the vehicle, totaling more than 250 thousand images. Figure \ref{fig:yolo_results} illustrates the results calibrated to remove detections of our own car. The results indicate that most objects detected are cars, traffic lights, trucks, and people. Having presented several dataset statistics relevant to its application, in the next section, we describe possible applications of the V2V dataset. 

\section{Enabled Applications} \label{sec:enabled_applications}
This section discusses the diverse applications enabled by the DeepSense 6G V2V dataset, spanning wireless communication, vehicular localization, and autonomous sensing applications. The multi-modal dataset provides invaluable resources for enhancing beamforming, predicting blockages, improving positioning systems, and developing efficient autonomous sensing algorithms. These applications highlight the wide-ranging impact of our dataset in advancing V2V communications and autonomous vehicle technologies.

\subsection{Wireless Communication Applications} \label{subsec:wireless_application}

This section presents two examples of V2V wireless communication applications enabled by multi-modal sensing provided as part of the DeepSense 6G V2V dataset.

\textbf{Beamforming and Beam Tracking:} To meet the high data rate demands of V2V communication, it is crucial to equip these systems with mmWave/THz transceivers, which require large antenna arrays and narrow directive beams to ensure sufficient signal-to-noise ratio. However, adjusting these narrow beams comes with a significant training overhead that scales with the number of antennas, posing challenges for supporting high-mobility V2V applications. Additionally, the highly mobile nature of V2V communication necessitates frequent updates to the optimal beam index that further increase this beam training overhead. The high mobility-induced frequent beam switching makes it difficult for these systems to meet future wireless communication application requirements, like low latency and high reliability. Delving deeper into the beam selection process reveals the following insights: Firstly, in mmWave/THz systems, beamforming is directional, which means that the optimal beam indices depend on the relative position of the transmitter and receiver. Secondly, objects in the wireless environment, whether stationary or moving, can affect the availability of the line-of-sight path and alter the optimal beam indices due to their limited multipath diversity and low penetration capability. Thirdly, the high mobility-induced latency can be minimized by enabling proactive decisions in the communication systems. Therefore, if the communication systems have access to information such as the location, mobility patterns, and geometry of the wireless environment, it may be possible to predict the optimal beams without relying on the conventional beam-sweeping method. These approaches are not limited to predicting the current optimal beams - they can be extended to predict future beams. 

This relevant information can be captured and extracted using additional sensors such as GPS receivers, cameras, LiDARs, and radars, making them promising candidates for enabling sensing-aided wireless communication applications. The DeepSense 6G V2V scenarios contain co-existing multi-modal data such as a $360$ camera, mmWave wireless communication, GPS data, 3D LiDAR, and radar collected in a real-wireless environment. The multi-modal nature of these scenarios helps enable several novel applications, such as sensing-aided multi-modal beam prediction and beam tracking and data fusion approaches for V2V communication systems. Combining data from different sensors may improve the accuracy and reliability of various V2V communication tasks. Moreover, the diversity of the V2V scenarios in the dataset, collected at different locations and times of the day, can help study the generalizability of the developed solutions. Generalizability is an important aspect of any machine learning or AI-based system, and the dataset diversity provides a valuable resource for assessing the robustness and adaptability of V2V communication solutions in different environments.

\textbf{Blockage Prediction and Beam Recovery:} The DeepSense V2V dataset can enable the development of algorithms for blockage prediction and beam recovery in wireless communication systems. The mmWave/THz communication systems rely on line-of-sight (LOS) links to achieve sufficient receive power. This is primarily due to the low penetration capabilities of the mmWave/THz signals, which makes LOS communication a dominant setting. Blocking these LOS links by either stationary or mobile objects in the environment can lead to significant degradation of the link quality and pose substantial challenges to the reliability and latency of these systems. Current approaches to link recovery are reactive, which incurs high latency in link re-connection, especially for mmWave/THz systems with very large codebooks and narrow directional beams. One way of enabling such proactiveness in wireless networks is by integrating and utilizing sensors such as GPS receivers, cameras, LiDARs, and radars to develop a comprehensive understanding of wireless environments. The additional information can help predict future blockages and initiate user handoff, thereby improving the reliability and latency of wireless communication systems. To achieve this, the DeepSense V2V dataset can be used to develop blockage prediction and beam recovery algorithms. The dataset provides data from multiple sensors, which can be integrated to develop a comprehensive understanding of the wireless environment. This approach can help initiate user handoff before a blockage occurs, reducing latency and improving the reliability of the system. In summary, the DeepSense V2V dataset provides a valuable opportunity to develop algorithms for blockage prediction and beam recovery in wireless communication systems. By integrating data from multiple sensors and developing a proactive approach, it is possible to predict future blockages and initiate user handoff beforehand, reducing the latency associated with link blockages and improving the reliability of wireless communication systems.

\subsection{Localization}
The DeepSense 6G V2V dataset includes data from multiple sensors, such as GPS, 3D LiDAR, radar, and vision sensors, which provide a comprehensive view of a vehicle's surroundings. This data can be used to develop and test vehicular positioning and navigation algorithms that can handle different driving scenarios and environmental conditions. Combining data from these different sensors makes it possible to develop algorithms that can accurately and reliably determine a vehicle's position and orientation. For example, GPS provides accurate location data, but its accuracy can be affected by signal interference and obstructions. Vision sensors and 3D LiDAR can provide more detailed information about the environment, such as the location and geometry of objects. This can help improve the accuracy and reliability of positioning and navigation. Moreover, the availability of multi-modal V2V data in the DeepSense 6G V2V dataset can help develop and test algorithms that can handle different driving scenarios and environmental conditions. For instance, vision sensors and 3D LiDAR can help provide more accurate and reliable location information in scenarios where GPS signals are weak or obstructed. Combining GPS, 3D LiDAR, radar, and vision sensors that provide 360-degree coverage can help achieve accurate and reliable vehicular positioning and navigation for V2V communication systems. The DeepSense 6G V2V dataset offers a valuable resource for developing and testing algorithms that can handle different driving scenarios and environmental conditions and improve the overall performance and robustness of V2V communication systems.

\subsection{Sensing Applications} \label{subsec:other_applications}

Apart from the wireless communication applications, the DeepSense 6G V2V dataset can be used to develop and test algorithms for various autonomous vehicle tasks. One such task is object detection and classification, which involves identifying and localizing different types of objects in the environment. Combining data from different sensors makes it possible to improve the accuracy and reliability of object detection and classification algorithms, which is critical for autonomous vehicles to navigate safely and efficiently. The 360-degree camera in the DeepSense 6G V2V dataset provides a comprehensive view of the environment, while the 3D LiDAR and radar sensors can provide detailed information about the location and geometry of objects in the environment. GPS data can also provide accurate location information, critical for object detection and classification. Moreover, the multi-modal nature of the DeepSense 6G V2V dataset can also enable the development of algorithms for other autonomous vehicle tasks, such as image segmentation, object tracking, and scene understanding. By leveraging the dataset multi-modal data, it is possible to improve the accuracy and reliability of object detection and classification algorithms, achieve more accurate and robust positioning and navigation, and develop algorithms for other AV tasks.

\section{Machine Learning Tasks} \label{sec:ML_tasks}
In machine learning, the development of practical solutions relies on several key components: a large-scale dataset, diversity in the data, access to ground truth labels, and the availability of comprehensive sensor information. These features collectively enable the development and evaluation of models that can generalize well and address real-world challenges. The DeepSense 6G V2V dataset offers a unique opportunity to explore and advance machine learning applications in the context of V2V communication. The multi-model sensing capabilities provide a comprehensive 360-degree view of the environment, and the incorporation of mmWave frequency arrays in the 60 GHz band makes the DeepSense 6G V2V dataset particularly significant for wireless communication research. Furthermore, the availability of different modalities permits modality fusion and allows for innovative solutions that leverage sensing and communication data integration.

Furthermore, the DeepSense 6G V2V dataset encompasses four distinct scenarios, each with its own characteristics and challenges. It consists of over 3.5 hours of data collected from various locations, time periods, and traffic conditions. This diversity reflects real-world complexities, enabling the development of models that can adapt to different environments and situations. The dataset includes intricate vehicle interactions, such as vehicles crossing each other or navigating multiple turns, presenting unique communication challenges. By incorporating these scenarios, the dataset facilitates the investigation and development of novel algorithms (such as sensing-aided beam and blockage prediction) that can handle real-world V2V communication challenges. The DeepSense 6G V2V dataset also benefits from a unified approach to data collection and structure across different scenarios. This unified framework ensures consistency and compatibility, which enables us to combine data from multiple scenarios to create larger development datasets. Advanced machine learning research avenues such as transfer learning, generalizability studies, scalability assessments, robustness evaluations, and distribution shift analysis can be explored by leveraging this capability. Moreover, the unified structure of the dataset enables the investigation of the generalization capabilities of machine learning models across different scenarios and the examination of the impact of distribution shifts on model performance. The DeepSense 6G V2V dataset enables innovative research in various machine-learning applications for V2V communication, and the following section explores a specific example: position-aided V2V beam prediction.

\subsection{Position-Aided V2V Beam Prediction} \label{subsec:position-beam}
Position-aided beam prediction utilizes GPS positions of vehicles to forecast the best beam index from a codebook, as demonstrated using the DeepSense 6G V2V dataset. This dataset includes precise position data for both transmitting and receiving vehicles, facilitating the development of algorithms that leverage this information to maximize received signal power. We aim to create a prediction solution that uses a sequence of position data points, not just a single pair, to enhance insight into the mobility and orientation of the vehicles involved in V2V communication. This sequence-based method offers advantages by providing a dynamic view of vehicle movement, including speed and acceleration, which helps predict trajectories more accurately. Moreover, understanding the orientation and movement of vehicles through sequential data is vital, especially when the receiver has multiple antenna panels, which adds complexity to beam prediction. This approach allows for more precise adaptations to various scenarios, such as rapid movements and complex interactions.

\subsection{Approach} \label{subsec:approach}

This section shows that this dataset makes position-aided beam prediction possible. One possible way of predicting the optimal beam using car positions is by engineering features that tightly correlate with the optimal beam index. We show that a sequence of positions can be used to determine the optimal beam by deriving the relative orientation and the relative positions between the two vehicles of each set of positions, applying a moving average across the sequence to smooth/average the noise and then using those positions to estimate the angle of arrival at the receiver vehicle (referred to as unit 1 in the previous sections). The previous statistics presented in Figures \ref{fig:beam_density} and \ref{fig:pos_density} from Section \ref{sec:Statistics} show that the relative position between the two vehicles and the beam index appears strongly correlated, suggesting this approach to be a good candidate to perform beam prediction. 

\begin{figure}[t]
	\centering
	\includegraphics[width=1\columnwidth]{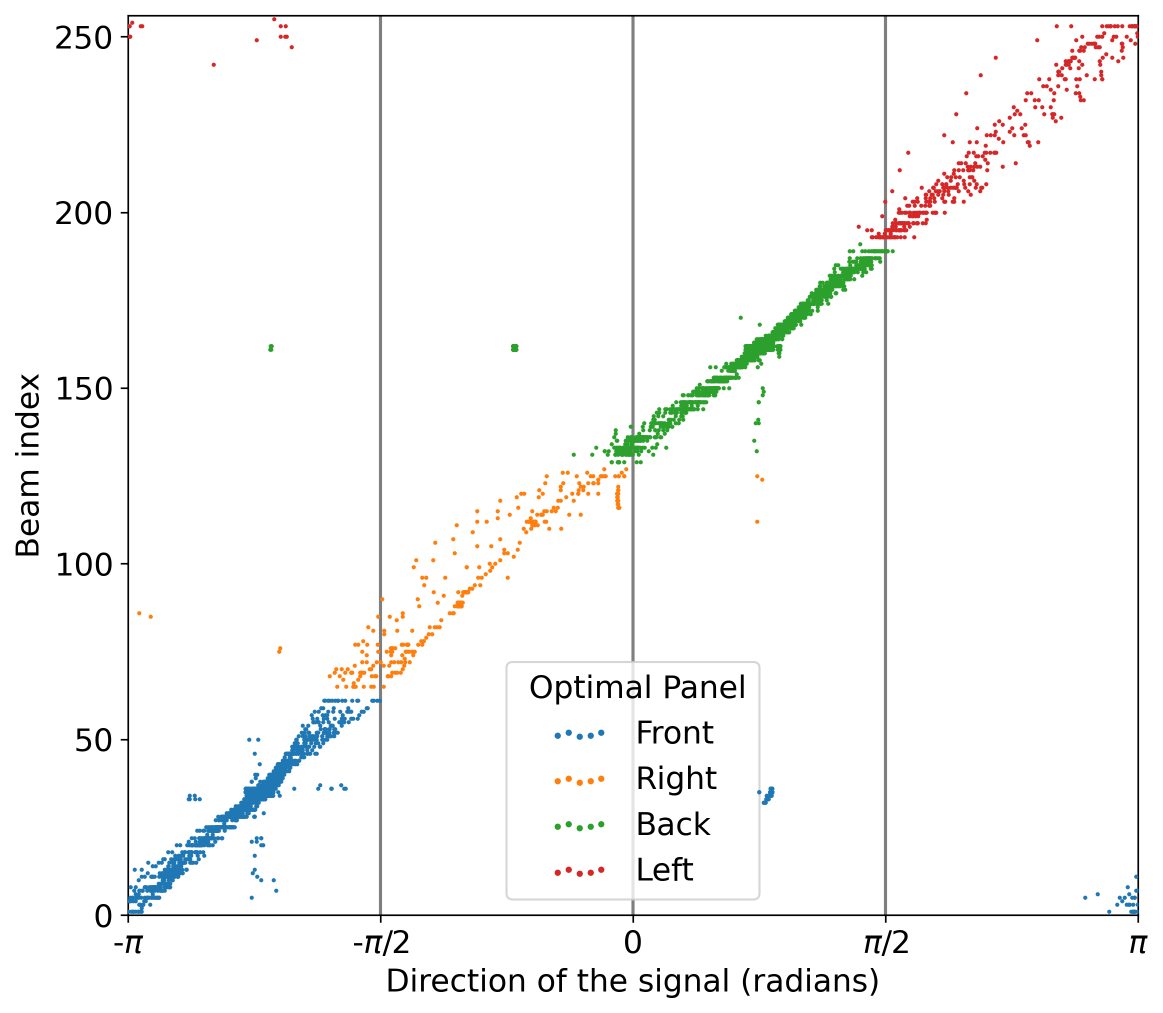}
	\caption{Correlation between the AoA estimated via GPS positions and the best beam index for Scenario 36. Vertical lines show the supposed panel separation according to the direction of the incoming signal, while colors show the ground truth optimal panel selection. When colors are outside their supposed interval, the optimal panel is not the expected panel, complicating optimal beam determination from the estimated AoA. }
	\label{fig:correlation}
        \vspace{-2mm}
\end{figure}

To estimate the angle of arrival in a predominantly single-path LoS setting, we need only the direction of the incoming wave with respect to the receiver and the orientation of the receiver. The direction of the wave can be estimated via the relative positions of the vehicles, and the orientation of the receiver can be similarly computed as the orientation of car/unit 1. Both quantities use ratios of latitudes and longitudes from the known formula of the angle of the slope

\begin{equation} \label{eq:base_formula}
    \theta\left(a, b\right) = \arctan{\left(\frac{\Delta_{lat}}{\Delta_{lon}}\right)}
\end{equation}

where $a$ and $b$ are the two positions necessary. Depending on the positions used in the formula, we either get the receiver orientation or the relative position of the two vehicles. Lacking better nomenclature, let $x_1 = (lat, lon)$ denote the position of vehicle 1 (receiver) and $x2$ be that of vehicle 2 (transmitter). If so, then the orientation of the receiver is given by $\theta\left(x_1(t), x_1(t-1)\right)$ and the relative position between receiver and transmitter is given by $\theta\left(x_1(t), x_2(t)\right)$. It should be noted that applying Equation \eqref{eq:base_formula} to compute these quantities still results in high sensitivity to noise. As such, we additionally apply a threshold filter (or high-pass) that considers $\Delta_{lat}$ or $\Delta_{lon}$ equal to zero whenever the difference is smaller than a certain quantity. In one expression, we write

\begin{equation}
    \Delta_{lat}(a,b) = 
    \left\{\begin{matrix}
    0 &\text{if} \quad \left |lat_a - lat_b \right| < lat_{thres} \\
    lat_a - lat_b &\text{if} \quad \left |lat_a - lat_b \right| > lat_{thres} \\
    \end{matrix}\right.
\end{equation}
and likewise for $\Delta_{lon}(a,b)$, with $lat_{thres} = lon_{thres} = 5e-7$ experimentally determined to be the smallest value that exceeded the GPS noise. The expression \eqref{eq:base_formula} is used twice for the arrival computation, as mentioned, but we first compute a simple moving average (SMA) on the estimates, i.e., an unweighted mean of the last $N_{avg}$ samples, to obtain better estimates that are more robust to noise. As such, mathematically we have
\begin{equation} \label{eq:moving_avg}
    SMA^{N_{avg}} (f, t) = \frac{1}{N_{avg}}\sum_{n_i=0}^{N_{avg}-1} f[t-n_i]
\end{equation}
with $N_{avg} = 30$ (i.e., average information from the last 3 seconds) and $f$ the function to be averaged. Finally, we use the smoothed estimates in the determination of the angle of arrival at time instant $t$
\begin{equation} \label{eq:aoa}
    AoA(t) = SMA(\theta(x_1(t), x_1(t-1))) - SMA(\theta(x_1(t), x_2(t))).
\end{equation}

\begin{figure}[t]
	\centering
	\includegraphics[width=1\columnwidth]{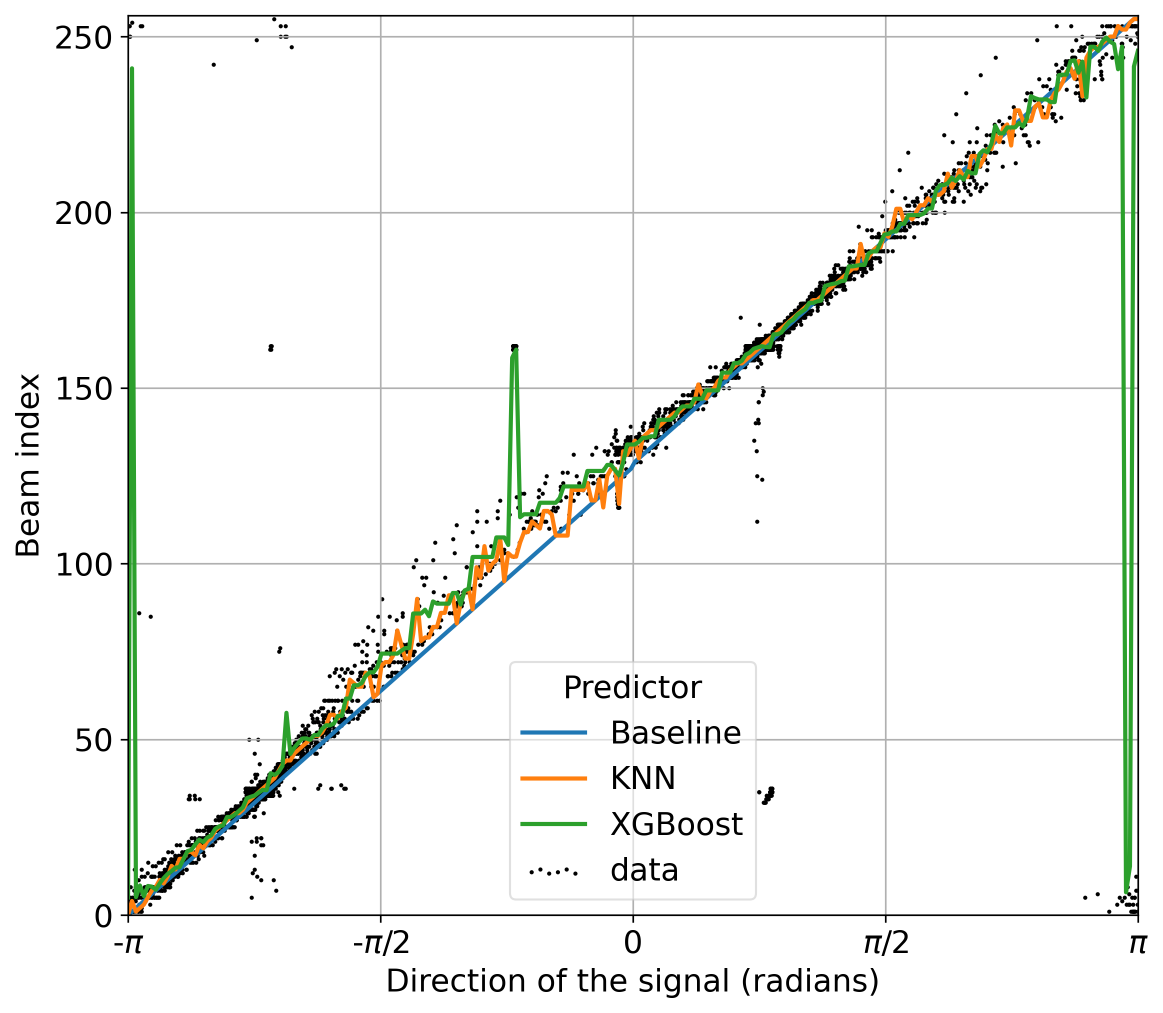}
	\caption{Fit from different linear and non-linear predictors to the AoA from GPS and beam index data from Scenario 36.}
	\label{fig:pred_fit}
\end{figure}

The process described here aims at maximizing the correlation of the $AoA$ and the optimal beam. We use the $AoA$ estimate in Equation \eqref{eq:aoa} and perform a mapping of the optimal beam indices to a uniform interval of $[-\pi, \pi]$. We display the relation between the two in Figure \ref{fig:correlation}. The figure shows a high correlation, suggesting that the $AoA$ will effectively determine the beam index. Like that, the problem is reduced to a regression where we try to select the mapping of $AoA$ to the beam index. To that end, we use several approaches. The first is a simple baseline, a uniform beam choice based on AoA that consists of an affine function of the form $y = m \phi$, with $\phi$ being the AoA in $[-\pi, \pi]$ radians and $y$ beams are uniformly distributed in azimuth from 0 to 255, then $m = 256 / (2 \pi)$. However, this heuristic is not resistant to real-world imperfections that cause data outliers, so better estimators should also be used.

The linear trend motivates other linear estimators, but they should be robust to noise and outliers. Literature shows us three linear estimators that are robust to noise: the Huber \cite{Huber}, the Ransac \cite{RANSAC}, and Theil-Sen \cite{Theil-Sen} estimators. We also consider non-linear estimators, such as KNN and the popular XGBoost \cite{XGBoost} for completeness. We fit these estimators to the data and show the baseline, KNN, and XGBoost results in Figure \ref{fig:pred_fit}. Because the remaining linear estimators have similar fits as the baseline, we omit them to make the figure less cluttered. Next, we look at how these estimators perform using classical performance metrics.

\subsection{Results}
\begin{figure*}[t]
	\centering
	\includegraphics[width=1.84\columnwidth]{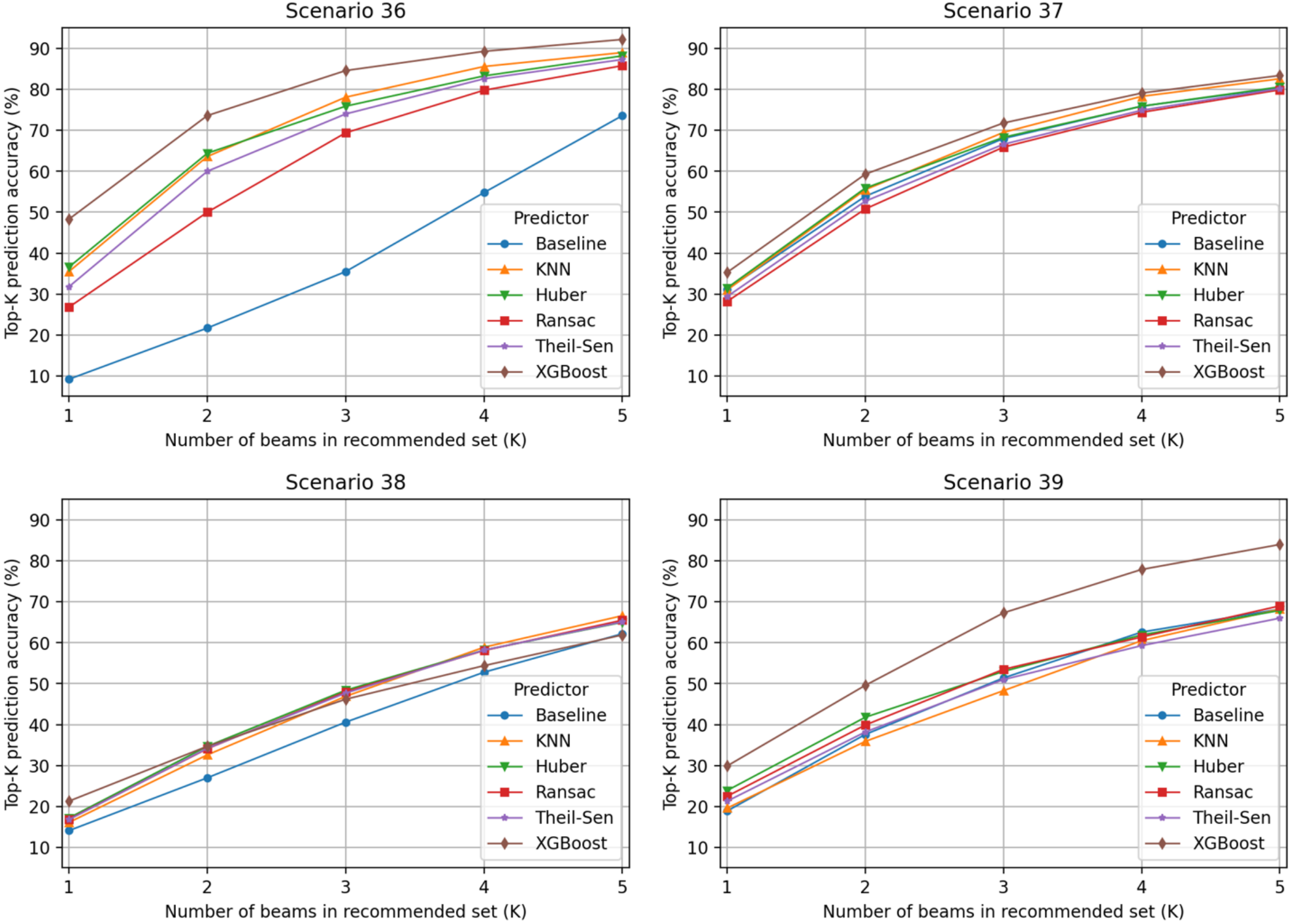}
	\caption{Regression results in top-k beam prediction accuracies from using different predictors in the estimation of AoA from GPS positions. }
	\label{fig:ml_results}
        \vspace{-2mm}
\end{figure*}

We used various methodologies for our regression analysis. These include the baseline (uniform heuristic), robust linear regressors (Huber, Ransac, and Theil-Sen), and non-linear estimators (KNN and XGBoost). The top-k accuracy curves are shown in Figure \ref{fig:ml_results}. These curves reveal a wide performance range across different scenarios. The top-5 accuracy varies significantly, between 60\% and 90\%. This variability is due to several factors. For example, the true signal angle of arrival (AoA) sometimes does not match the relative orientation derived from GPS positions. This mismatch mainly occurs when buildings or other obstacles cause non-line-of-sight (NLoS) signal propagation. Additionally, even when the receiver and transmitter are completely still, hardware noise can cause changes in the chosen beams. To mitigate this effect, we filtered our data with a signal-to-noise ratio (SNR) threshold of 0 dB. This is because when the SNR is less than 0 dB, any beam can be chosen, and that choice may not correlate with the positions. Finally, position noise can vary between different scenarios. This variation can affect our AoA estimation. Considering the performance of different estimators, XGBoost appears to be the most effective. However, the simpler KNN estimator achieved similar results. This finding was surprising, as KNN performed as well as more complex estimators known for their robustness to outliers.

\section{Conclusion}
This work presents DeepSense V2V, the vehicle-to-vehicle scenarios of the DeepSense6G dataset. We provided an in-depth exploration of the dataset, illustrating its creation process and potential applications in the interplay of communications, sensing, and localization. We began by detailing the DeepSense6G scenario creation pipeline, which encompasses data collection, processing, and visualization. Subsequently, we demonstrated the diversity of the dataset by offering comprehensive statistics on various road types and locations, vehicle velocities, beam distributions, and road-related object detection. As a practical example, we utilized the dataset to predict beam directions based on GPS positions. We expect this dataset to serve as a significant asset for research in both academia and industry, enhancing studies in wireless communications and advancing autonomous driving technologies.

\bibliographystyle{IEEEtran}

\end{document}

%% file: input.tex
\usepackage{amsfonts}
\usepackage{times}
\usepackage{graphicx}
\usepackage{latexsym}
\usepackage{dsfont}
\usepackage{amssymb}
\usepackage{amsmath}
\usepackage{cite}
\usepackage{verbatim}
\usepackage{subfigure}

\def\bb0{{\mathbb{0}}}

\def\bb{{\mathbf{b}}}

\def\b0{{\mathbf{0}}}

\def\sf0{{\mathsf{0}}}

